\documentclass[num-refs]{nbdt-article}

\usepackage{siunitx}
\usepackage{graphicx}

\papertype{Original Article}
\paperfield{Journal Section}

\title{Mixed-horizon optimal feedback control as a model of human movement}

\author[1]{Justinas Česonis}
\author[1,2,3]{David W. Franklin}

\affil[1]{Neuromuscular Diagnostics, Department of Sport and Health Sciences, Technical University of Munich, Munich, Germany}
\affil[2]{Munich Institute of Robotics and Machine Intelligence (MIRMI), Technical University of Munich, Germany}
\affil[3]{Munich Data Science Institute (MDSI), Technical University of Munich, Munich, Germany}

\corraddress{Department of Sport and Health Sciences, Technical University of Munich, Munich, 80992, Germany}
\corremail{david.franklin@tum.de}

\runningauthor{Česonis and Franklin}

\begin{document}

\maketitle

\begin{abstract}
Computational optimal feedback control (OFC) models in the sensorimotor control literature span a vast range of different implementations.
Among the popular algorithms, finite-horizon, receding-horizon or infinite-horizon linear-quadratic regulators (LQR) have been broadly used to model human reaching movements.
While these different implementations have their unique merits, all three have limitations in simulating the temporal evolution of visuomotor feedback responses.
Here we propose a novel approach -- a mixed-horizon OFC -- by combining the strengths of the traditional finite-horizon and the infinite-horizon controllers to address their individual limitations.
Specifically, we use the infinite-horizon OFC to generate durations of the movements, which are then fed into the finite-horizon controller to generate control gains.
We then demonstrate the stability of our model by performing extensive sensitivity analysis of both re-optimisation and different cost functions.
Finally, we use our model to provide a fresh look to previously published studies by reinforcing the previous results, providing alternative explanations to previous studies, or generating new predictive results for prior experiments.
\\

\end{abstract}

\section{Introduction}

Computational modelling has driven our understanding of human sensorimotor control by supplementing experimental results and motivating new hypotheses \cite{Hogan1984a, Flash1985a, Harris1998, Uno1989a}.
In particular, inspiration from control engineering theory has recently contributed to new ideas of how humans plan and execute movements \cite{Scott2004c}.
For example, robust control inspired models have been considered in order to guarantee stability in the presence of noise \cite{Franklin2008c, Crevecoeur2013b, ueyama_mini-max_2014, crevecoeur_robust_2019, bian_model-free_2020}. 
Similarly, optimality principles have been proposed to explain human movement and solve issues of redundancy through trade-offs between different elements of the cost function, for example task goals and energy consumption \cite{Todorov2002c, Todorov2005, Liu2007, Shadmehr2016}.
Overall these computational approaches have been very successful at reproducing and explaining human-like behaviours \cite{Izawa2008c, Guigon2007, Nagengast2009a, Yeo2016}. 
Among numerous studies, different optimal control algorithms are applied to seemingly similar experimental paradigms.
However, there are subtle differences in implementations of different optimal control paradigms that could result in meaningful behavioural differences, so the motivation of using any one specific algorithm is not always clear.

The majority of the control algorithms used for simulating optimal feedback control strategies rely on the linear-quadratic regulator (LQR) or linear-quadratic Gaussian (LQG) techniques.
However, distinctions between the types of LQG or LQR implementations, such as the infinite, finite or receding time horizon, are still underdiscussed, which is particularly important as these different implementations result in different simulated behaviours.
For example, in goal directed reaching movements, the finite-horizon control will accurately predict human-like dynamics such as feedback responses to a perturbation \cite{Liu2007, Cesonis2020}, as well as some kinematics, like undershooting the target.
However, in many experimental studies, where participants are required to finish each trial within the target, this undershooting also becomes an inconsistency, and a model limitation.
In contrast, similar simulations with infinite or receding horizon models will produce appropriate kinematics \cite{Jiang2011, Guigon2019}, but will fail to accurately model the feedback responses \cite{Cesonis2020}.
Here, we propose that in order to simulate more realistic movements using these paradigms we should utilise the strengths of multiple different implementations, rather than weighing the pros and cons of each algorithm to minimise the drawbacks.

Previous research has shown that humans plan and execute movements in separate steps, which likely occur in different brain areas {\cite{Cisek2010, Kodl2011}}.
As such, it is not unreasonable to assume that these steps could be modelled by different algorithmic implementations.
Particularly, here we propose a new, mixed-horizon approach in modelling movement planning and execution where the planning stage is represented by an infinite-horizon optimal feedback controller, and the execution stage is represented by a finite-horizon OFC, with the infinite-horizon controller providing movement durations to the finite-horizon controller.
While we often consider planning as occurring prior to the movement initiation, here the mixed-horizon model uses the infinite-horizon controller to re-plan the movement duration (or time-to-target) after any perturbation throughout the movement. 
That is, the planning and execution processes continue throughout the entire movement, allowing the model to respond to any unseen or unpredicted perturbations with appropriate human-like changes.
As a result, the combined system allows us to benefit from the individual strengths of the two controllers while addressing each of their limitations.

\section{Materials and Methods}

\subsection{Mixed-horizon optimal feedback controller}

In this article we propose a new optimal feedback control (OFC) framework for modelling human reaching movements, called the mixed-horizon OFC.
We have termed this framework as mixed-horizon, as it combines the features of finite-horizon control \cite{Todorov2005, Todorov2005b}, and infinite-horizon control \cite{Jiang2011, Qian2013}.
Specifically, we have previously shown \cite{Cesonis2020} that even though the infinite-horizon controllers produce reaching movements of an appropriate duration even after the movement is visually perturbed, the corrections to these perturbations do not vary in intensity in the same way as they do in human movements \cite{Cesonis2020, Franklin2008, OostwoudWijdenes2011, Dimitriou2013, Franklin2016}.
On the other hand, the finite-horizon controller produces feedback responses with variable intensity that depend on the time-to-target \cite{Cesonis2020}, consistent with the data of human participants, but inherently requiring movement duration as an input variable.
By combining the two frameworks into the single mixed-horizon controller we can overcome the limitations of each individual model and generate a more human-like control response.
For completeness, we present two different types of models: a simplistic, separable model, implemented as a linear-quadratic regulator (LQR) and assuming a perfect observer, and a more advanced non-separable model implemented as iterative linear-quadratic Gaussian (iLQG) and with signal-dependent noise present.

A particular limitation of finite-horizon implementations for the modelling of perturbed goal directed movements is that they inherently require movement duration (or time-to-target) as an input variable.
However, in most real-life and experimental cases this movement duration is not predefined.
While it is possible to estimate this duration from the data, or set experimentally for non-perturbed movements, we have previously demonstrated that human participants non-trivially extend their movement times post-perturbation if the goal of the task is to reach the target \cite{Cesonis2020}.
Thus, for every perturbed movement, unless all movement variables (time, distance, perturbation onset, perturbation magnitude) perfectly match, the movement duration would need to be separately estimated in order to accurately apply the finite-horizon OFC.
For tasks where model fitting is of interest such limitations may be addressed by individually assessing different types of movements within the task, however such implementation would still not generalise to unseen perturbations.

The limitation of the required input duration for the finite-horizon models can technically be addressed in a multitude of different ways: arbitrary choice, use-dependence, temporal discounting of reward \cite{Shadmehr2016}, feed-forward learning, or feedback control (particularly infinite \cite{Jiang2011, Qian2013} or receding horizon \cite{Guigon2019}).
However, while there could definitely be feedforward effects in non-perturbed baseline movements, previous research indicates that feedback processes play a significant role in setting the movement duration post-perturbation \cite{Cesonis2020, Georgopoulos1981}, as this duration is adjusted based on perturbation onset, type or magnitude.
On the other hand, both infinite and receding horizon OFC implementations have been shown to reliably predict movement durations for both non-perturbed and perturbed movements, independent of the perturbation onset, and without changing control parameters. 
Therefore, there exists an infinite (or receding) horizon OFC with a fixed set of control gains, that could immediately and reliably compute the movement duration for the non-perturbed movement, and in case of a perturbation at any point during this movement -- immediately recompute an appropriate, extended movement duration.
In turn this duration can then be used to adjust the finite-horizon control policies post-perturbation.
Finally, while we previously have shown that the infinite-horizon OFC and receding-horizon OFC can both predict the post-perturbation movement durations equally well, for this paper we utilised the infinite-horizon architecture due to fewer model parameters (i.e. no required length of horizon as an additional input).

\subsection{Experimental data}
The focus of this article is on building the computational model using the OFC framework.
However, to illustrate the usability of our model, we apply it to model the results of previous behavioural studies.
Specifically, we model the findings of goal directed reaching movement studies with cursor and/or target perturbations \cite{Cesonis2020, Dimitriou2013, Franklin2016}.
We chose these particular studies, as their experimental findings show the regulation of feedback responses to cursor or target perturbations that traditional OFC models could not replicate.
Moreover these studies demonstrate complex modulation of feedback gains across a variety of time points, perturbation magnitudes, and conditions that we can use to test our models.

\subsection{State space representation}
For all models we used the same state space representation.
The hand was modelled as a point mass with $m = 2$ kg.
The intrinsic muscle damping was modelled as viscosity $b = 10$ Ns/m (consistent with \cite{Liu2007} for movements of comparable speed).
This point mass was controlled in a two dimensional x-y plane by two orthogonal force actuators that simulated muscles.
These actuators were regulated by a control signal $u_t$ via a first-order low-pass filter with a time constant {$\tau = 0.06$ s}.
The generic state-space representation at time $t$, used to simulate the system, could be written as

\begin{equation} \label{met:eqn1}
	x_{t+1} = Ax_t + Bu_t + \epsilon_tBCu_t + \xi_t,
\end{equation}

\noindent where $A$ is a state transition matrix (in discrete time, here only shown for one spatial dimension):

$$
A = \begin{bmatrix}
1 & \delta t & 0\\
0 & 1-b \delta t/m & \delta t/m\\
0 & 0 & 1-\delta t/\tau
\end{bmatrix}
$$

\noindent and $B$ is a control matrix (in discrete time, here only shown for one spatial dimension):

$$
B = \begin{bmatrix}
0\\
\delta t/\tau\\
0
\end{bmatrix}
$$

\noindent $\xi_t$ represents additive control noise, which for our simulation purposes was always set to zero.
$\epsilon_tC$ represents signal-dependent multiplicative noise where $\epsilon_t$ is sampled at time $t$ from zero mean and unit variance Gaussian noise, and $C$ is a 2 $\times$ 2 scaling matrix defining the magnitude of this noise.
The numeric values of $C$ were adjusted to simulate low, medium or high noise levels for non-separable models, or set to zero for the separable model.
State $x_t$ exists in the Cartesian plane and consists of position $\vec{p}$, velocity $\vec{v}$ and force $\vec{f}$ (two dimensions each). 
For our model implementations in discrete time we used the sampling rate $\delta t = 0.01$ s

The state of the plant that is used for control, $x_t$, is not directly observable, but has to be estimated from the system's output, $y_t$ via the output equation 

\begin{equation} \label{met:eqn2}
	y_{t} = Hx_t + \sigma_tD.
\end{equation}

\noindent For our models we set $H = I_6 = diag(1,1,1,1,1,1)$, meaning that all our hidden state variables (position, velocity, force) are observable by the controller which is consistent with human physiology. 
$\sigma_t$ is sampled at time $t$ from zero-mean unit variance Gaussian noise and $D$ is a diagonal matrix representing the intensity of this noise on all output states. 
In the non-separable models we set $D = 0.015\, diag(1,1,1,1,10,10)$, consistent with values in \cite{Liu2007}.
Even though the same values could be used, in the separable models, for simplicity we set $D$ = 0.

\begin{figure}[t!]
	\centering
	\includegraphics[width=1\linewidth]{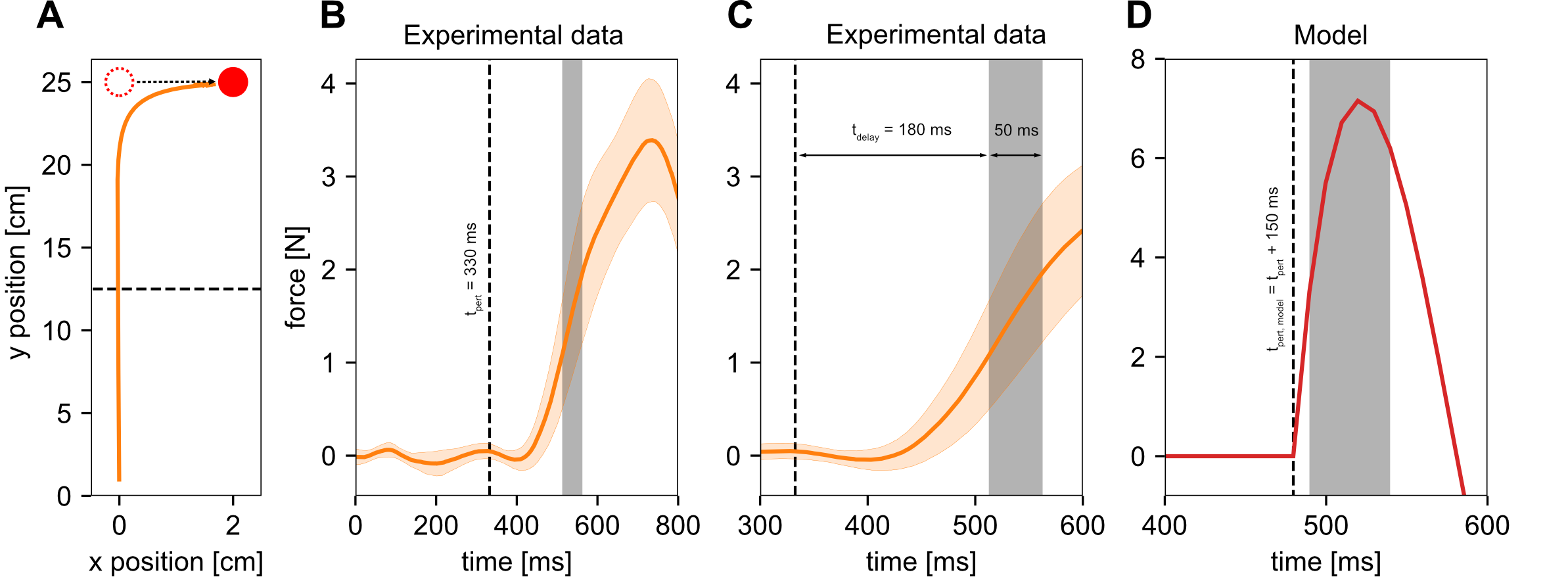}
	\caption{An example of visuomotor feedback responses and intensities. \textbf{A.} A typical movement trajectory to the visual perturbation of the target, perpendicular to the direction of movement. The perturbation is induced as a target jump from its original position (dashed red circle) to the new position (solid red circle) when the movement trajectory passes the perturbation onset location. An example here shows perturbation induced at the mid-distance along the movement. Note that x and y axes are not shown to scale for better visibility, and participants do not undershoot the target in experimental data. \textbf{B.} A typical motor response to the visual perturbation of the target, perpendicular to the direction of movement. No force responses are present in the perturbation direction until the perturbation is induced. An exemplar perturbation (same as in \textbf{A}) at t = 330 ms from the beginning of the movement produces a force response in the direction of the perturbation approximately 150 ms later, due to visuomotor delays. \textbf{C.} Same data as in \textbf{B}, zoomed to time of the perturbation and the response. Force response is averaged over the time window of [180 ms - 230 ms] from the onset of the perturbation (grey shaded area) to determine the response intensity. \textbf{D.} Equivalent perturbation induced in the model simulations. Due to no delays in the computational model, the perturbation is induced 150 ms later compared to the human participants. In addition, due to the faster ramp-up of the model responses, the feedback intensity is computed by averaging the response over [10 ms - 60 ms] after the perturbation (gray shaded area).}
	\label{fig:methods}
	
\end{figure}

\subsection{Estimation of feedback responses}
We apply all our models described in this paper to simulate visuomotor feedback responses \cite{Sarlegna2003, Saunders2003, Brenner2003, Franklin2008} by mimicking the target jump or cursor jump paradigms.
In these studies, a visual perturbation of either a target or a cursor during a movement induces a feedback motor response that allows participants to bring the cursor to the target (Figure \ref{fig:methods}A).
When measured in terms of corrective force, this response is typically delayed about 150 ms from this perturbation onset until the corrective force is produced (Figure \ref{fig:methods}B).
In order to quantify the magnitude of this response, the force is typically averaged over a time window of [180 - 230] ms after the perturbation \cite{Franklin2008, Dimitriou2013, Franklin2016, Cesonis2020} to produce a visuomotor feedback intensity (Figure \ref{fig:methods}C).
This intensity has previously been shown to vary based on the perturbation onset location \cite{Dimitriou2013}, magnitude \cite{Franklin2016}, time-to-target \cite{Cesonis2020} or task dependency \cite{Franklin2008}.
Note that in many previous studies the feedback intensity was referred to as the feedback gain, however as we use the term gain to describe the control gains in this article, we opted to use feedback intensities to define these averaged force responses.

In order to estimate the appropriate kinematic behaviour and simulated feedback responses we first trigger the perturbations based on the forward movement of the cursor.
Specifically, based on the experimental design of the simulated study there exists a perturbation onset location along the forward movement.
Once the cursor crosses this location, the perturbation timer of {150 ms} is started to simulate the visuomotor feedback delay.
After this delay, the perturbation is triggered by shifting the cursor or the target appropriately to the task design (Figure \ref{fig:methods}D).
Thus, from the external observer's perspective any perturbations in our model are happening 150 ms later than in human experiments.
However, considering the delays present in human visuomotor system, the motor responses in human participants would be observed at the similar time as in our model, for the perturbations with the same onset.

In our models in this study we induced the visuomotor delays to the feedback system of our model by simply delaying the perturbation by the fixed duration.
This results in the similar behaviour between the model and human participants, as the responses are produced at the matching times.
An alternative implementation for the delay could be a fixed intrinsic system delay.
However, this implementation would require expanding the state space by 15 times (with our model sampling rate of 100 Hz and 150 ms visuomotor delay), and adding additional computational load. 
Our preliminary work showed little differences in these two implementations for simulation of these visuomotor feedback responses.

After the perturbation is physically induced we record the average response of the model to the perturbation over the time window of [10 - 60] ms to calculate the feedback intensity. 
This is equivalent to [160 - 210] ms time window in humans which is somewhat shorter than the conventional [180 - 230] ms time window. 
However, this is necessary as our models produce faster responses than humans and is also consistent with our previous work \cite{Cesonis2020}.

Finally, while there is a distinction between responses to cursor perturbations and target perturbations in human participants \cite{Reichenbach2014}, in terms of the modelling there is no functional distinction between the target or cursor jumps, as our models only account for the difference vector between them to estimate an appropriate control signal.
Therefore, the selection of the perturbed entity (cursor or target) has no effect on our final simulation results.

\subsection{Separable mixed-horizon OFC}
Separable mixed-horizon OFC is a simplified model with no additive or multiplicative control noise and perfect state information. 
This model consists of two major building blocks: the infinite-horizon controller, implemented as an infinite-horizon linear quadratic regulator (LQR), and a finite-horizon LQR controller.
First, we use the infinite-horizon LQR to simulate the movement from start to target position in order to estimate the appropriate movement duration for such a movement. 
This could be considered as a movement planning stage.
Second, we use that movement duration as a parameter for the simulation using the finite-horizon controller.
As a result, the two parts are connected linearly and can thus be implemented separately and then combined together.

Infinite-horizon LQR is a simplistic optimal controller that generates an optimal control solution to a given system with state cost $x^T_tQx_t$ and control cost $u^T_tRu_t$.
We define the generic form of the performance index $J$ as

\begin{equation} \label{met:eqn3}
	J = \sum^{\infty}_{t=0} x^T_tQx_t + u^T_tRu_t = \sum^{\infty}_{t=0} \omega_p(\vec{p}_t - \vec{p^*})^2 + \omega_v||\vec{v}_t||^2 + \omega_f||\vec{f}_t||^2 + \omega_r||u_t||^2
\end{equation}

\noindent where $\omega_p$, $\omega_v$ and $\omega_f$ are position, velocity and force state cost parameters, $\omega_r$ is the activation cost parameter and $\vec{p^*}$ is a target position.
For the finite-horizon controller this performance index is instead

\begin{equation} \label{met:eqn4}
	J = \sum^{N}_{t=0} x^T_tQ_tx_t + u^T_tR_tu_t = \sum^{N}_{t=0} \omega_{p, t}(\vec{p}_t - \vec{p^*})^2 + \omega_{v, t}||\vec{v}_t||^2 + \omega_{f, t}||\vec{f}_t||^2 + \omega_{r, t}||u_t||^2
\end{equation}

\noindent where $N$ is the duration of the movement obtained via the infinite-horizon controller. 
Note here that the control parameters for the finite horizon can be non-stationary and thus be different for every time-point.
As a general rule unless stated otherwise, for fast, goal-directed reaching movements we set $Q = 0$ for $t \neq N$, and kept $R$ stationary, consistent with \cite{Todorov2005, Liu2007}.

\subsubsection{Model optimisation}

For a given mechanical system, the behaviour of the LQR system is defined by the control parameters $\omega_p$, $\omega_v$, $\omega_f$ and $\omega_r$. 
Thus, instead of selecting the values of control parameters arbitrarily (e.g. from previous literature or by qualitative inspection of final system kinematics or dynamics) we opted to quantitatively optimise our model for the best suited parameters.
We executed the optimisation for the infinite-horizon part and the finite-horizon part separately due to their linear relation.

Within our mixed-horizon implementation we use the infinite-horizon part to estimate the durations of individual movements.
As a result, we optimise the control parameters of the infinite-horizon part based on the fit between the experimental movement durations, and the movement durations produced by our model in an equivalent paradigm.
Specifically, we optimise these parameters on the mean durations of all available perturbed movements simultaneously for the study that we are modelling.
Thus, with a single set of optimal control parameters, the infinite-horizon controller produces the movement duration that matches the movement duration of any experimentally perturbed movement, when this matching perturbation is induced.
In this way, instead of searching for a new set of control parameters (and thus deriving new controller gains) for every different perturbation, we ensure that for any perturbation our infinite-horizon controller always produces a movement duration that matches that of the human participants, without changing control parameters or control gains.
We used the sum of squared-residuals as a goodness of fit measure between the real experimental movement durations and those produced by our models, and the Nelder-Mead algorithm to find the best fit.
Furthermore, in order to reduce the chance of finding a local minimum we instantiated the optimisation {10} times with random initial conditions and selected the solution with the lowest sum of square-residuals (SSR).

We use the finite-horizon implementation in order to obtain the kinematics and dynamics for each model.
As a result, we aim to fit the kinematics of the finite-horizon simulations to the kinematics of the movement from experimental data.
In order to evaluate how well the model kinematics fit to the experimental kinematics we devised a kinematic cost function

\begin{equation} \label{met:eqn5}
	\begin{split}
		\Gamma &= \gamma_1 (v_{peak,\,req} - v_{peak,\,model})^2 + \gamma_2 (p_{peak,\,req} - p_{peak,\,model})^2  \\
		&+\gamma_3 (v_{end,\,req})^2+\gamma_4 (p_{end,\,req} - p_{end,\,model})^2 
	\end{split}
\end{equation}

\noindent where the four components are squared-errors of the peak velocity magnitude, forward position of where the peak velocity was achieved, the end-point velocity and the end position (i.e. the distance of the under/overshoot).
The parameters $\gamma_{1-4}$ are the relative weights assigned to each of the components.
The $v_{peak,\,req}$, $p_{peak,\,req}$ and $p_{end,\,req}$ are fixed quantities dependent on the experiment of interest and are measured in cm/s and cm respectively.
Equation \ref{met:eqn5} is designed to specifically simulate the experimental conditions that participants in the experiments were instructed to follow.
Instead of fitting the whole kinematics, we optimise the kinematics to the similar ``instructions'' as the participants are given.
For example, if participants are instructed to “stop at the target” (meaning have zero velocity and zero error from the target), “try to produce movements that are the right speed” (indirectly instructing peak velocity via “too fast” and “too slow” feedback), or in some specific cases (i.e. experiments in study 1) providing a direct feedback of peak velocity location, our kinematic cost function provides similar constraints to our model via parameters $\gamma_1 - \gamma_4$.

\subsubsection{Model sensitivity to the kinematic cost function}
The kinematic cost function influences the desired model behaviour.
For example, based on Equation \ref{met:eqn5}, a model where $\gamma_1 >> \gamma_2, \gamma_3, \gamma_4$ would reward kinematics that match the peak velocity requirement, while placing less emphasis on the peak velocity location, final position or final velocity.
In contrast, a model with $\gamma_3 >> \gamma_1, \gamma_2, \gamma_4$ would prefer kinematics with no residual velocity while relaxing demands on the other three components.
As a result, we conducted a sensitivity analysis on the kinematic cost function to estimate a range of different model behaviours when performing the same task.
To do so, we first selected a range of different kinematic cost functions.
For the baseline $\Gamma$ we used relative weights ($\gamma_1, \gamma_2, \gamma_3, \gamma_4$) = (4, 4, 0.25, 25).
For all other $\Gamma$ we kept three of the relative weights at their baseline value, while varying the fourth one.
We chose $\gamma_1 \in [0.125, 16]$, $\gamma_2 \in [0.125, 16]$, $\gamma_3 \in [0.125, 16]$ and $\gamma_4 \in [0.25, 400]$.
For each $\Gamma$ we then performed a full model optimisation to find the control parameters $\omega_p$, $\omega_v$, $\omega_f$ and $\omega_r$ that minimise this given $\Gamma$.
As previously, we conducted a Nelder-Mead optimisation with {10} different instances and selected a solution with the minimum $\Gamma$ as a single output.
Finally, we analysed the resultant model dynamics qualitatively by comparing the simulated force responses (intensities) to the visuomotor feedback intensities in the experimental data.

\subsubsection{Model sensitivity to individual optimisations}
We also tested how sensitive our model is to individual optimisation instances. 
Specifically, for every individual optimisation the best fit behaviour is achieved with a different set of parameters ($\omega_p$, $\omega_v$, $\omega_f$, $\omega_r$).
Thus, we also analysed the relative distribution of these best fit parameters of individual optimisations, and how this change influences the dynamics and kinematics of the control system.
We performed optimisation sensitivity analysis {40} times on the baseline kinematic cost function.

\subsection{Non-separable mixed-horizon OFC}
The implementation of non-separable mixed-horizon OFC is similar to the separable mixed-horizon OFC.
The main difference between the two is the presence of multiplicative control noise in the finite and infinite-horizon blocks.
For the infinite-horizon part, instead of using an infinite-horizon LQR implementation we adapted the implementation used by \cite{Qian2013}, where the control noise was transformed to a loss on control gains and iteratively optimised until convergence.
For the finite-horizon part of the model we used the iterative LQG algorithm to obtain the control policy in presence of noise \cite{Todorov2005, Liu2007}.

In presence of multiplicative control noise both infinite and finite-horizon LQG become computationally more expensive due to the iteration until convergence when calculating control and observer gains.
As a result, to make the optimisation manageable, for the models with multiplicative noise we initiated every individual optimisation {three times instead of the 10 }that we used for the separable models, and selected the best value.
While this increased the chance that the performance of the model is sub-optimal, our models still behaved generally similar to the separable conditions.

\subsubsection{Noise parameters}
In total we tested the optimisation and controller behaviour in three different noise conditions.
We introduced these conditions via a noise scaling matrix $C$ (Equation \ref{met:eqn1}).
{The scaling matrix was defined as

\begin{equation} \label{met:eqn6}
	C = k\begin{bmatrix}
		0.15 & 0.05 \\
		-0.05 & 0.15
	\end{bmatrix}
\end{equation}
}
\noindent with $k$ = 1 for low noise condition, $k$ = 3 for medium noise condition and $k$ = 5 for high noise condition.
The implementation of $C$, where noise is proportional not only on the activation of the actuator responsible for moving in a desired direction, but also to the activation of a perpendicular one, induces a coupling between the two Cartesian actuators, consistent with human muscles.
The values of $k$ were selected by trial and error: $k=1$ produces a control behaviour which is generally stable, but different from no-noise behaviour, and $k>5$ often results in optimisation timing out without finding a minimum.
In comparison, a typical human participant from \cite{Cesonis2020} produced motor variability that is comparable to our model at $k=5$, but ranged between $k=3$ and $k=10$ for different participants.
As the human participants experience contributions from other variability sources (e.g. planning variability, signal independent noise, measurement noise) in addition to the control-dependent noise, the actual level of control-dependent noise is therefore likely comparable to our selected values.
Finally, while this does not strictly mean that $k=5$ mimics high motor noise, with our implementation we could not consistently test any higher noise values.

\subsection{Specific model implementations}
In order to demonstrate the versatility of the mixed-horizon model, we used our model to replicate the experimental behaviour results of previous studies.
As different studies have slight differences in implementations, our model had to be adapted for an individual study to comply with the design.
While the methods listed above are common across different studies that are modelled in this article, the current section details specific differences for each study.

\subsubsection{Study 1: Česonis \& Franklin 2020 \cite{Cesonis2020}}

In the original study participants were asked to reach to the target while producing a specific velocity profile: baseline, early peak (movement accelerated early and slowed for the latter portion), or late peak (velocity gradually increased and reached the peak late in the movement) \cite{Cesonis2020}.
From the modelling perspective, these three different conditions were implemented in the finite-horizon OFC by making the activation cost $\omega_r$ time-dependent:

\begin{equation} \label{met:eqn7}
	R'(t)= \frac{\exp (p \frac{t+q}{r})}{mean(R')}
\end{equation}
\vspace{-30 px}
\begin{equation} \label{met:eqn8}
	\omega_r = \omega_r R'(t)
\end{equation}

Here $p$, $q$ and $r$ are the three parameters governing the temporal shape of the activation cost function.
In order to maintain activation cost of an equivalent magnitude between conditions, the $R'$ was normalised to the mean value of 1.

For the finite-horizon part of our model we implemented this scaling to produce the local offset of the peak velocity.
However, the same approach can not be applied to the infinite-horizon part of the model. 
Particularly, the temporal evolution of the $\omega_r$ assumes the known movement duration, so that it can then be normalised over the movement duration. 
If this duration is not known, then this temporal profile becomes undefined.
As a result, the infinite-horizon part did not modulate peak velocity locations, but instead we re-fit the control parameters to adapt the movement duration to the experimental condition. 

While selecting a new set of controller parameters seems unreasonable for modelling human-like systems, we utilised the model (in)sensitivity to different optimisations to find a parameter space, such that we can switch between conditions within the infinite-horizon part by only changing one of $\omega_p$, $\omega_v$, $\omega_f$ and $\omega_r$. 
Such adaptation is more realistic than changing all parameters -- it is reasonable to assume that different required movement kinematics would impart a change in some, but not necessarily all, control parameters.

{In order to evaluate the model performance in modelling study 1, we simulated feedback responses equivalent to those in the experiment.
In the experiment, as the participants reached towards a target located 25 cm away from the start position, their cursor representation was occasionally perturbed perpendicularly to the movement direction by 2 cm.
These perturbations were implemented as cursor jumps either to the left or to the right, and induced as the hand crossed one of five onset locations (P1, P2, P3, P4, P5), evenly spaced along the movement distance (4.2, 8.3, 12.5, 16.7, 20.8 cm).}
Similarly, our model simulated each of these perturbation conditions -- five different perturbation onset locations for each of the three velocity conditions -- and generated a total of 15 responses.

\subsubsection{Study 2: Dimitriou et al. 2013 \cite{Dimitriou2013}}
Experiment 1 in \cite{Dimitriou2013} is conceptually very similar to the study described above.
As a result, here we focus on simulating the results of experiments 2 and 3.
In total, for each experiment we simulated four different perturbed movements.
Two of these movements were to a stationary target: cursor-perturbed reaching to a "far" target (25 cm) and cursor-perturbed reaching to a "near" target (17.5 cm), where the cursor was perturbed laterally by 2 cm.
In the other two movements, in addition to these cursor perturbations the target was also perturbed in the movement direction: in movements where the starting target was "near", the target was perturbed to "far" and vice-versa.
In all conditions of experiment 2 both cursor and target perturbations were induced simultaneously when the participant had moved 15.75 cm from the start position.
In experiment 3 the target perturbations were induced at 10.5 cm, instead of 15.75 cm, and the cursor perturbations were still induced at 15.75 cm. 
In experimental data this resulted in about a 100 ms time difference between the two perturbations \cite{Dimitriou2013}.

The major difference between study 1 and study 2 is that in  study 2, perturbations only happened in channel trials (there were no maintained perturbations as in study 1).
As a result, we modified our model implementation to simulate these channel trials by fixing the x-position (position perpendicular to the start-target axis) to 0 when the cursor is not perturbed, and to 2 cm when the cursor is perturbed. 
We only constrained this position in the simulation of the movement, but not during the planning of the movement (i.e. when the controller is being calculated).

In terms of the baseline movements (no perturbations) study 2 has similar requirements to study 1. 
As a result, for the finite-horizon part of our model we used the same model parameters as in the study 1 baseline condition. 
However, we have tuned the parameters of the infinite-horizon part of the model so that durations of simulated movements better fit the durations of the movements in the experimental data. 
Here for all the 4 types of movement and both experiments we used the same set of parameters.

\subsubsection{Study 3: Franklin et al. 2016 \cite{Franklin2016}}
In this study participants were asked to reach to a target 25 cm away from the start position with some trials being perturbed during the movement.
The perturbations could happen either to the target, the cursor, or both, and were always lateral to the movement direction.
In terms of magnitude, perturbations could be 0 cm, 1 cm, 2 cm or 3 cm for both target and cursor, and could occur to either left or right.
As one of the findings, the authors showed that for isolated perturbations (only cursor or only target perturbations) the response intensity increased with perturbation magnitude in a non-linear, saturating manner {(see Figure 1D,E in \cite{Franklin2016}).}
We simulated these isolated perturbations in our model in order to first replicate, but also gain further insight into possible mechanisms governing this phenomenon.

In the original study, the lateral cursor or target perturbation always occurred when participants had moved 12.5 cm from the start position. 
However, in a simulated experiment we induced these perturbations at five equally spaced locations along the movement (one of them being at 12.5 cm).
With more simulated perturbations we can analyse the behaviour of the controller not only at the original (experimental) perturbation, but also around it in order to see whether this non-linear modulation arises from the general regulation of the response, or if it can be explained by the time-to-target regulation \cite{Cesonis2020}.
In the former case, we would expect to see stronger overall responses with increasing perturbation size, independent of the onset position/time.
In the latter case, we would expect to see comparable responses across the conditions as long as the time-to-target matched.

Similar to our model of study 2, we maintained the same finite-horizon parameters as the baseline model of our study 1, as the requirements for movements are similar.
However, we again fit the infinite-horizon parameters to better match the movement durations between the model and the data.

\subsection{Code implementation}
All code was written in Python 3.7.1 in a Spyder 4.1.3 environment.
Array computations and optimisations were executed with \texttt{NumPy v1.19.0} and \texttt{SciPy v1.5.1} libraries. 
Where necessary, state-space systems were discretised from continuous time using a zero-order hold method available in \texttt{control v0.8.3}.

\subsection{Code availability}
The code for all the simulations that are described in the paper, as well as the data obtained via these simulations is freely available online at \url{https://doi.org/10.6084/m9.figshare.14356346}.

\section{Sensitivity analysis}
We performed sensitivity analysis on our model before selecting the model parameters for any given simulation of an experiment.
For this analysis we used the mixed-horizon model implementation that we built to simulate our previous results \cite{Cesonis2020}.
In order to understand how sensitive the model behaviour is to individual optimisations with the same kinematic cost function (Equation \ref{met:eqn5}), we first performed optimisation sensitivity analysis.
This is necessary due to the fact that our optimisation minimum is not a single point within the parameter state-space, but a set of points.
Each set of points could potentially result in different model behaviour, as each optimisation run produces a different set of model parameters.
However, such a model would not be reliable for our applications, so this has to be tested first.
On the other hand, such analysis might allow us to uncover an underlying structure of the optimal parameter space, if multiple parameter sets produce identical optimal behaviours.
Second, we performed sensitivity analysis on different kinematic cost functions that we use to optimise the model behaviour.
While it is expected that changing the kinematic cost function will affect the model behaviour in general, our purpose with this analysis is to see how sensitive the relative behaviour is between the conditions (e.g. the relative regulation of simulated responses in three different kinematics of our previous work \cite{Cesonis2020}).
Specifically, by analysing the cost sensitivity we can test whether particular outcomes of the model simulations result from model fitting, or whether that behaviour arises from the structure within the model.
This distinction is important, as the former would indicate that a vast range of behaviours could be fit to a specific model, indicating that the model structure is not meaningful.
However, the latter would indicate that the behaviour itself is not an outcome of the parameter choice, but results from the model structure, and thus the behaviour is robust.

\subsection{Optimisation sensitivity}
The mixed-horizon OFC consists of two main parts -- the infinite-horizon and the finite-horizon OFC working in series. 
As each part is responsible for different functions within the whole model, namely, determining the movement duration, or, computing the optimal kinematics and dynamics for that duration, the two parts can be optimised separately.
In this section we present the optimisation sensitivity analysis for both parts of separable (without control noise) and non-separable (with control noise) models.

\begin{figure}[p!]
	\centering
	\includegraphics[width=\linewidth]{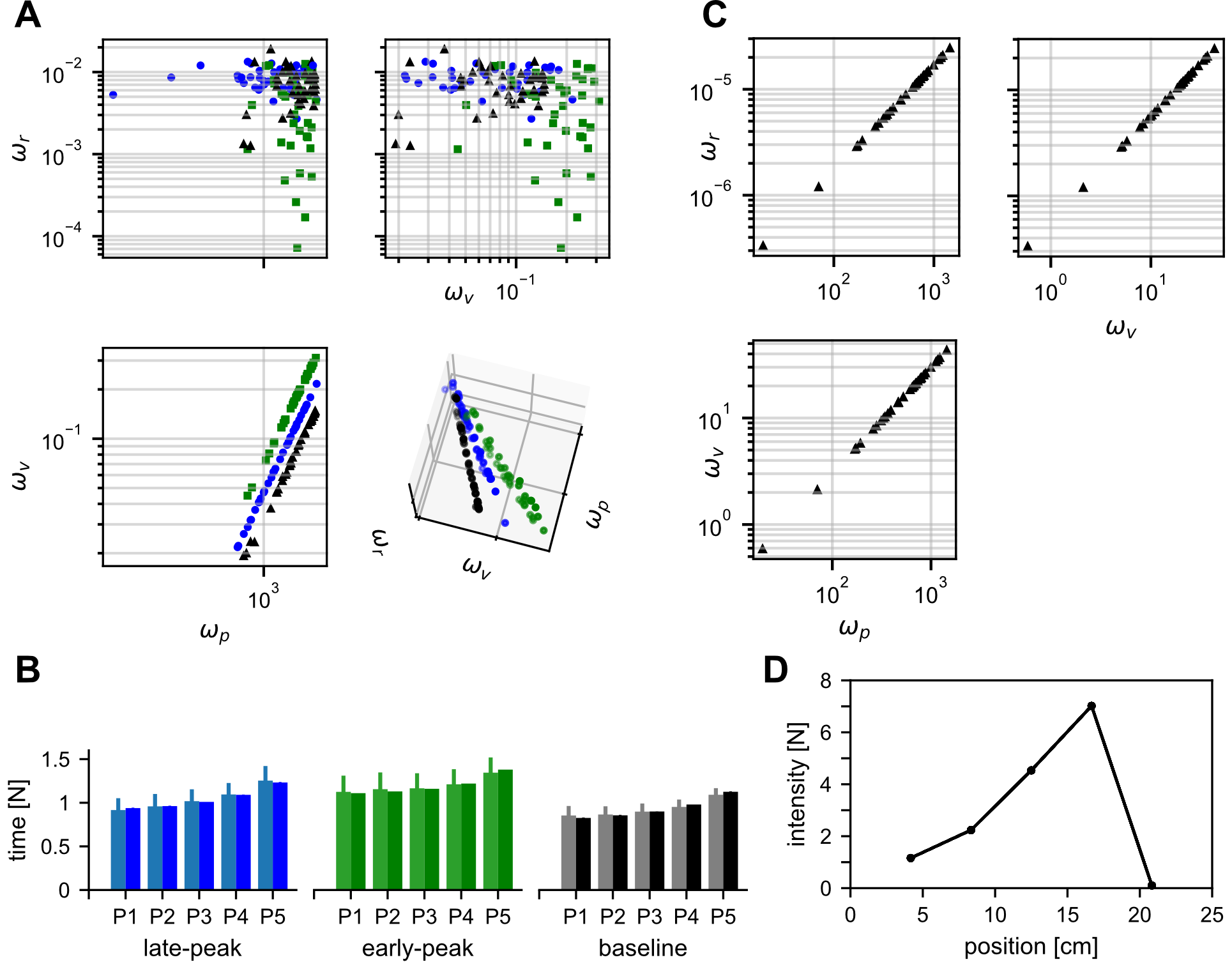}
	\caption{Optimisation sensitivity analysis for the separable model. \textbf{A.} Distribution of the best LQR infinite-horizon control parameters in the $\omega_p$-$\omega_v$-$\omega_r$ parameter space. Each dot represents one set of parameters, while separate plots show the same parameters in a different projection. The fourth control parameter $\omega_f$ is not shown, as it is tied to $\omega_v$ in a fixed relationship. Black triangles represent the baseline condition, green squares the early-peak velocity condition, blue circles the late-peak velocity condition simulations. Optimal parameters from different conditions do not overlap in $\omega_p-\omega_v$ projection except at the origin, but do in the other two projections. \textbf{B.} Movement duration sensitivity to different optimisation instances. Blue (left), green (middle) and black (right) bars represent the movement durations of the late-peak, early-peak and baseline conditions. Lighter shaded bars (left bars for each perturbation) show the mean and 95\% confidence intervals (95\% CI) of the experimental movement durations for each condition and each of five perturbations (P1-P5). Darker shaded bars (right bars for each perturbation) show the movement durations (and 95\% CI) simulated by our separable infinite-horizon model. \textbf{C.} Distribution of the best LQR finite-horizon control parameters in the $\omega_p$-$\omega_v$-$\omega_r$ parameter space for the baseline condition. Each dot represents one set of parameters, while separate plots show the same parameters in a different projection. The fourth control parameter $\omega_f$ is not shown, as it is tied to $\omega_v$ in a fixed relationship. Optimal parameters are systematically distributed on a line in the parameter space. Early-peak and late-peak conditions are not shown, as they share the same parameter values (with additional parameters to determine the shape of the velocity profile). \textbf{D.} Finite-horizon LQR simulated feedback intensities for the baseline condition. Each set of parameters (one dot) from \textbf{C} is used to generate one profile (trace) of simulated feedback intensities by simulating five perturbed movements with different perturbation onset locations. Each of five dots here represents the perturbation onset position, and the corresponding response intensity. All {40} of simulated intensity profiles overlap demonstrating consistency across optimisations.}
	\label{fig:sep_opt_sen}
	
\end{figure}

\subsubsection{Separable infinite-horizon controller}
An infinite-horizon controller in the mixed-horizon framework is used to estimate the total movement duration, which is then passed to the finite-horizon controller to generate movement.
In order to fit this controller to produce movement durations, we optimised its state and activation costs $\omega_p$, $\omega_v$, $\omega_f$ and $\omega_r$ so that the movement durations matched those of human participants in equivalent conditions \cite{Cesonis2020}.
Specifically, for each of the three kinematics (baseline, early-peak velocity and late-peak velocity) we simulated a movement where the cursor was perturbed in each of the five perturbation onset locations (once per movement) and evaluated the goodness of fit by calculating the sum of square residuals (SSR) between the generated and actual movement durations.
We used this goodness of fit, together with Nelder-Mead optimisation, to find the $\omega_p$, $\omega_v$, $\omega_f$ and $\omega_r$ that provide the best fit between model and data durations.
We repeated this optimisation {40 times} for each condition, generating a set of control parameters per optimisation.
These parameters are shown in Figure \ref{fig:sep_opt_sen}A.

In order to evaluate the sensitivity of the simulated movement durations to different optimisation runs we computed the 95\% confidence intervals (95\%CI) for these simulated durations and compared them to the equivalent 95\%CI for human movements.
Despite the wide variations in model parameters for each optimisation, our 95\%CI for simulated data is well within the respective intervals for human data, meaning that our simulations are less variable than human movements.
Thus, these results show that the model is not sensitive to different instances of the optimisation.

\subsubsection{Separable finite-horizon controller}

Similar to the infinite-horizon part of the model, we analysed the sensitivity of the finite-horizon controller to different runs of optimisation.
While for the infinite-horizon part we evaluated the goodness of fit during the optimisation via the SSR between movement durations, here we already use the movement duration as the model input.
Instead, for the finite-horizon part we minimise the kinematic cost function $\Gamma$ (Equation \ref{met:eqn5}) in order to produce movements that are kinematically similar to those of the human participants.

The distribution of the optimal state and activation costs over different optimisations is shown in Figure \ref{fig:sep_opt_sen}C.
While every optimisation yielded a different solution numerically, all of these solutions were distributed on the same line in the parameter state-space. 
In addition, no differences were observed across simulations in either the kinematics or dynamics.
Moreover, each simulation produced a consistent variation of the feedback intensities across perturbation locations (Figure \ref{fig:sep_opt_sen}D).
Thus, all together this shows that the separable finite-horizon OFC is extremely robust to different instances of optimisation.

\begin{figure}[t!]
	\centering
	\includegraphics[width=\linewidth]{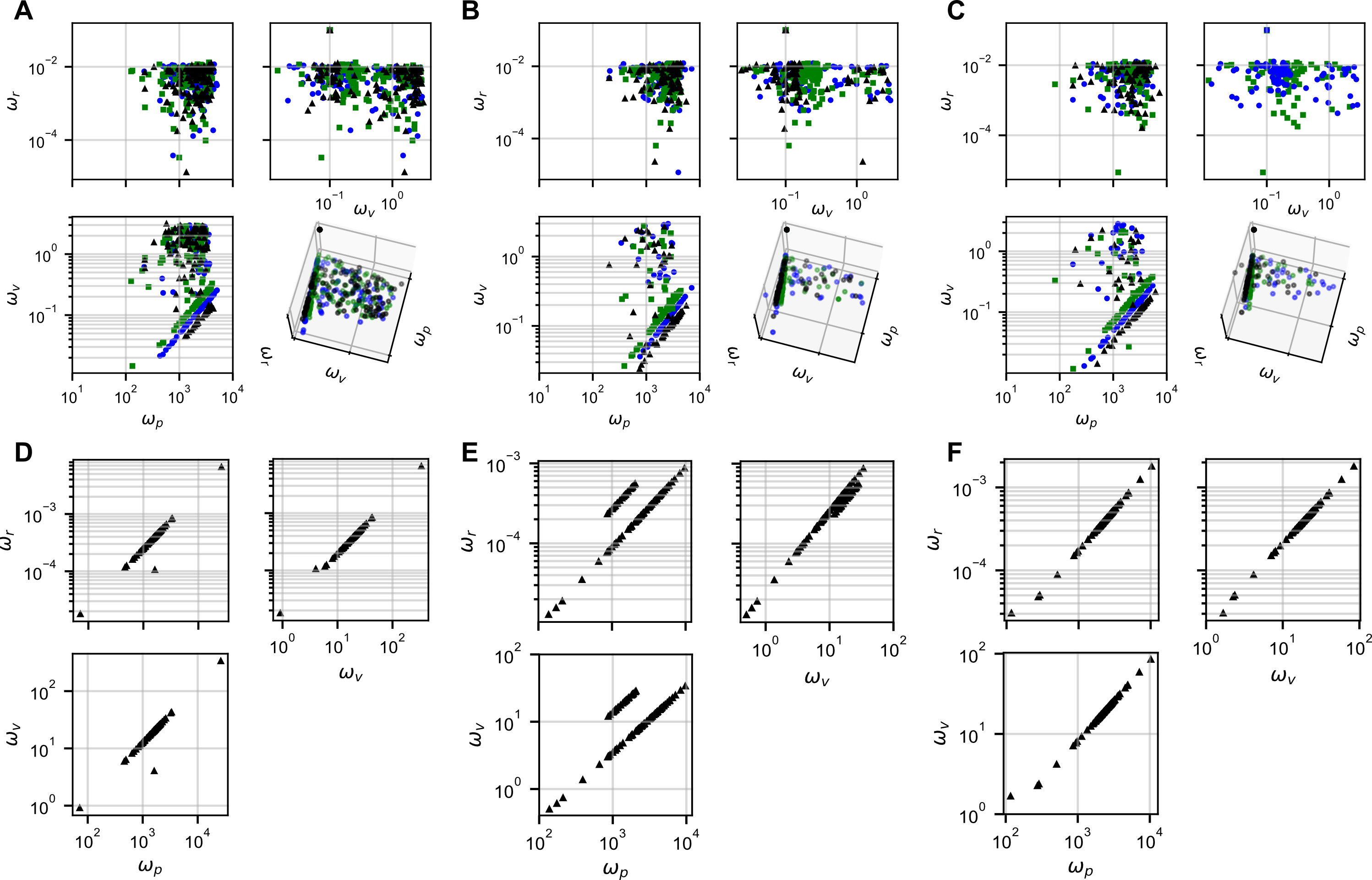}
	\caption{Optimisation sensitivity analysis for the non-separable model. \textbf{A.} Distribution of the best LQG infinite-horizon control parameters in the $\omega_p$-$\omega_v$-$\omega_r$ parameter space for noise level $k = 1$. Each dot represents one set of parameters, while separate plots show the same parameters in a different projection. The fourth control parameter $\omega_f$ is not shown, as it is tied to $\omega_v$ in a fixed relationship. Black triangles represent the baseline condition, green squares the early-peak velocity condition, blue circles the late-peak velocity condition simulations. Successful optimisations converge onto the parameter plane perpendicular to the $\omega_p-\omega_v$ plane and do not overlap across the conditions. Non-successful optimisation outputs are also shown even though they did not converge, and produce undesirable behaviour. \textbf{B, C.} Same as in \textbf{A,} but for noise levels $k = 3$ and $k = 5$ respectively. \textbf{D.} Distribution of the best LQG finite-horizon control parameters in the $\omega_p$-$\omega_v$-$\omega_r$ parameter space for noise level $k = 1$ in the baseline condition. Successful optimisations converged to a line in the parameter space. Unsuccessful optimisations deviate from this line, and are not shown as they are mainly outside of figure boundaries. Early-peak and late-peak conditions are not shown, as they share the same parameter values (with additional parameters to determine the shape of the velocity profile). \textbf{E, F.} Same as in \textbf{D,} but for noise levels $k = 3$ and $k = 5$ respectively. Interestingly, the optimal parameter space for $k = 3$ results in two lines in the parameter space.}
	\label{fig:nonsep_opt}
\end{figure}

\subsubsection{Non-separable infinite-horizon OFC}
Here we analysed the sensitivity of the non-separable infinite-horizon OFC to different optimisation runs. 
While the non-separable model is similar to the separable model in its implementation, a key difference is the noise in the system which influences model behaviour and leads to a different optimal control solution.
Similar to the separable model, the non-separable infinite-horizon OFC is used to compute the required movement duration that would be used by the finite-horizon part.
As a result, the sensitivity analysis for the non-separable model is largely similar to the separable model -- we used the SSR between the model predicted movement duration and the experimental movement duration as a goodness of fit measure.

In total we simulated the non-separable models with three different levels of noise (Equation \ref{met:eqn6}) $k=1$, $k=3$ and $k=5$.
For each of the three noise levels we initially ran the optimisation {120 times per kinematic condition (360 optimisations total per noise level)}.
For each of these optimisations we observed one of three different outcomes:
\begin{enumerate}
	\item Optimisation converged to a global minimum, with parameters producing consistent and replicable movement durations;
	\item Optimisation converged to a local minimum due to random initial conditions, with parameters producing inconsistent movement durations;
	\item Optimisation did not converge, or converged with parameters that are out of bounds (e.g. negative costs).
\end{enumerate}
For each noise level we selected only the parameters that converged to a reliable solution (outcome 1).
Specifically, with each set of obtained ($\omega_p$, $\omega_v$, $\omega_f$ and $\omega_r$) we simulated the movement durations via the infinite-horizon controller three times, and only selected those sets of parameters where the {geometric mean of the sums of squared-errors between simulated durations and experimental durations was less than $2.15 \times 10^{-3}$ s$^2$.
This method helped filter out solutions that belong to option 2, however occasional outliers were still present after such filtering.
As a result, we performed an additional outlier removal step using the $1.5 IQR$ (interquartile range) rule, applied on mean simulated movement durations.
After the outlier removal, there remained 132 parameters sets for $k=1$ , 182 parameter sets for for $k=3$ and 177 parameter sets for $k=5$.}
Finally, in order to keep the number of reliable solutions for $k=1$ comparable to the other two noise conditions, we ran an additional 60 optimisations per kinematic condition (180 total optimisations) for this noise condition.
This brought the total of remaining parameter sets for $k=1$ noise condition to {206}.

\begin{figure}[t!]
	\centering
	\includegraphics[width=\linewidth]{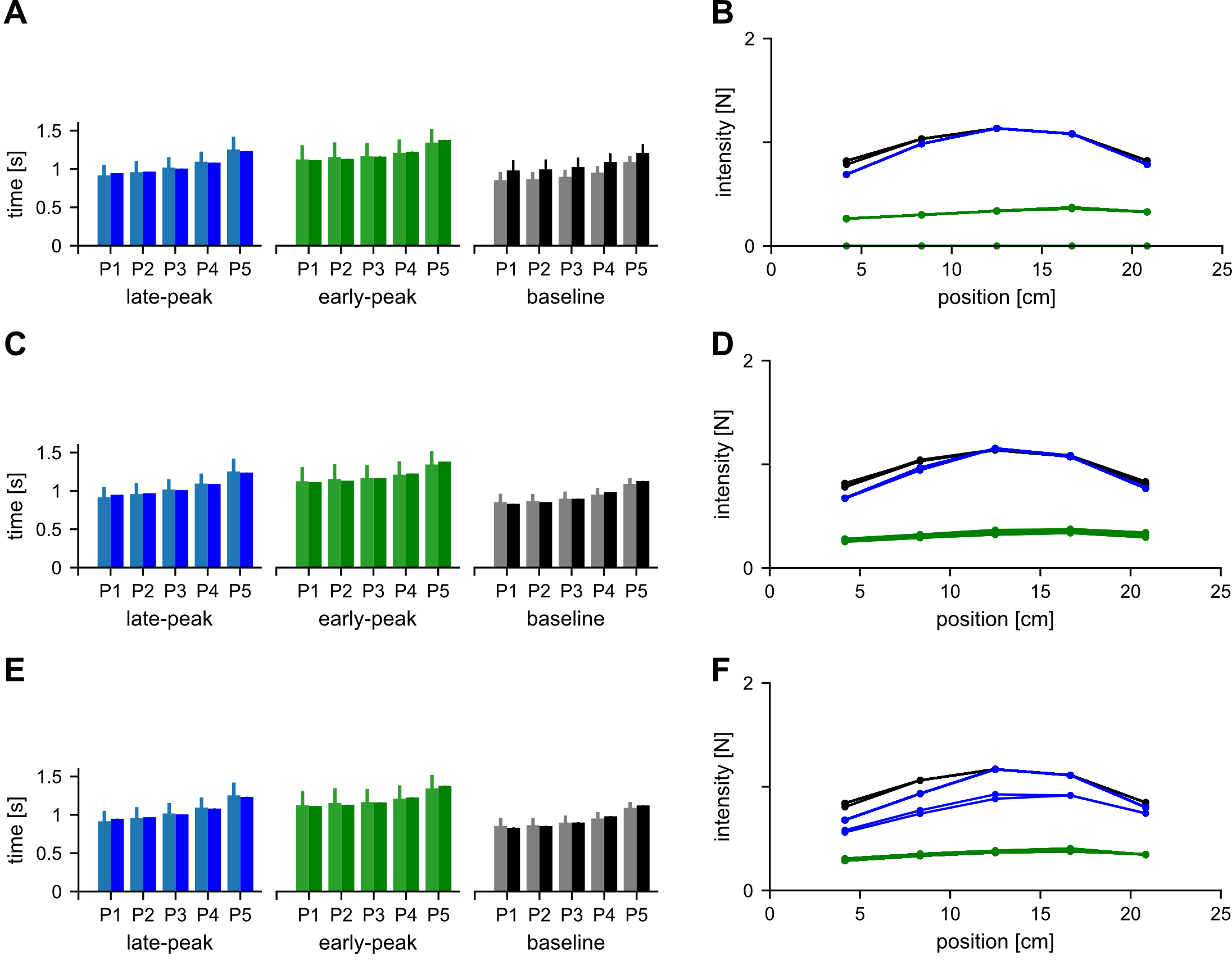}
	
	\caption{Sensitivity of the non-separable mixed-horizon OFC to different optimisation instances. \textbf{A.} Blue (left), green (middle) and black (right) bars represent the late-peak, early-peak and baseline movement durations for noise level $k=1$. Lighter shaded bars (left bars for each perturbation) show the mean and 95\% CI of the experimental movement durations for each condition and each of five perturbations (P1-P5). Darker shaded bars (right bars for each perturbation) show the movement durations simulated by our non-separable infinite-horizon model. \textbf{B.} Simulated feedback intensity sensitivity to different optimisation instances using the mixed-horizon OFC for noise level $k = 1$.
	Each line trace represents a feedback intensity profile, generated by simulating five perturbed movements with different perturbation onsets, but with the same control parameters.
	Twelve simulated profiles are shown for the baseline condition (black lines) along with sixty profiles each for the early peak (green) and the late-peak (blue) conditions.
	Only convergent optimisation outputs are used. The converging outputs produce consistent behaviour even when the final parameters are different. \textbf{C, D.} Same as \textbf{A, B.} but noise level $k=3$. \textbf{E, F.} Same as \textbf{A, B.} but noise level $k=5$. In \textbf{F.} 58 out of 60 simulation profiles overlap, while two simulations produced lower intensities.}
	\label{fig:nonsep_timegain}
\end{figure}

The resulting optimal parameters for all three noise levels are visualised in Figure \ref{fig:nonsep_opt}A-C.
From all successful simulations we also estimated the mean and 95\%CI of movement durations for each noise level (Figure \ref{fig:nonsep_timegain}A,C,E).
{Except for $k = 1$ baseline condition, all model simulations show lower or comparable variation than the human data, suggesting that the model is sufficiently insensitive to different optimisations.
For the $k=1$ baseline condition, the model mean is shifted towards high values due to a few outliers. 
While these outliers generally worsen our results, they can easily be spotted qualitatively and thus removed, not compromising model behaviour when performing simulations.}

\subsubsection{Non-separable finite-horizon OFC}
As previously, we optimised non-separable finite OFC by evaluating the goodness of fit via a kinematic cost function $\Gamma$ (Equation \ref{met:eqn5}).
For each of the three noise levels $k=1$, $k=3$ and $k=5$ we fit our model to the baseline condition {200 times}.
As in the infinite-horizon optimisations, some of these optimisations did not converge to a stable solution (e.g. by producing negative parameter values).
In total, {there were 89, 107 and 114 successful optimisations} for $k=1$, $k=3$ and $k=5$ respectively, shown in Figure \ref{fig:nonsep_opt}D-F.

We then randomly took {12} sets of our baseline optimisation output parameters, and used them to find best-fit values $p$, $q$ and $r$ (Equation \ref{met:eqn7}) for generating early-peak and late-peak velocity conditions of the movement.
For each set of these 12 baseline parameters we repeated the process {5} times per non-baseline condition to generate a total of {120} sets of ($p$, $q$, $r$) parameters (60 per non-baseline condition).
Although these parameters differed numerically, the resulting behaviour was similar (Figure \ref{fig:nonsep_timegain}B,D,F).

\subsubsection{Mixed-horizon OFC sensitivity to different optimisations} 
Overall our results of model sensitivity show that the model is reasonably robust to different optimisations.
Depending on the noise level in the system the optimisation may time-out without converging to the minimum.
However, when the optimisation successfully converged, even though the resultant parameters were numerically different, they produced largely similar behaviour within our model.

\begin{figure}[p!]
	\centering
	\includegraphics[width=0.99\linewidth]{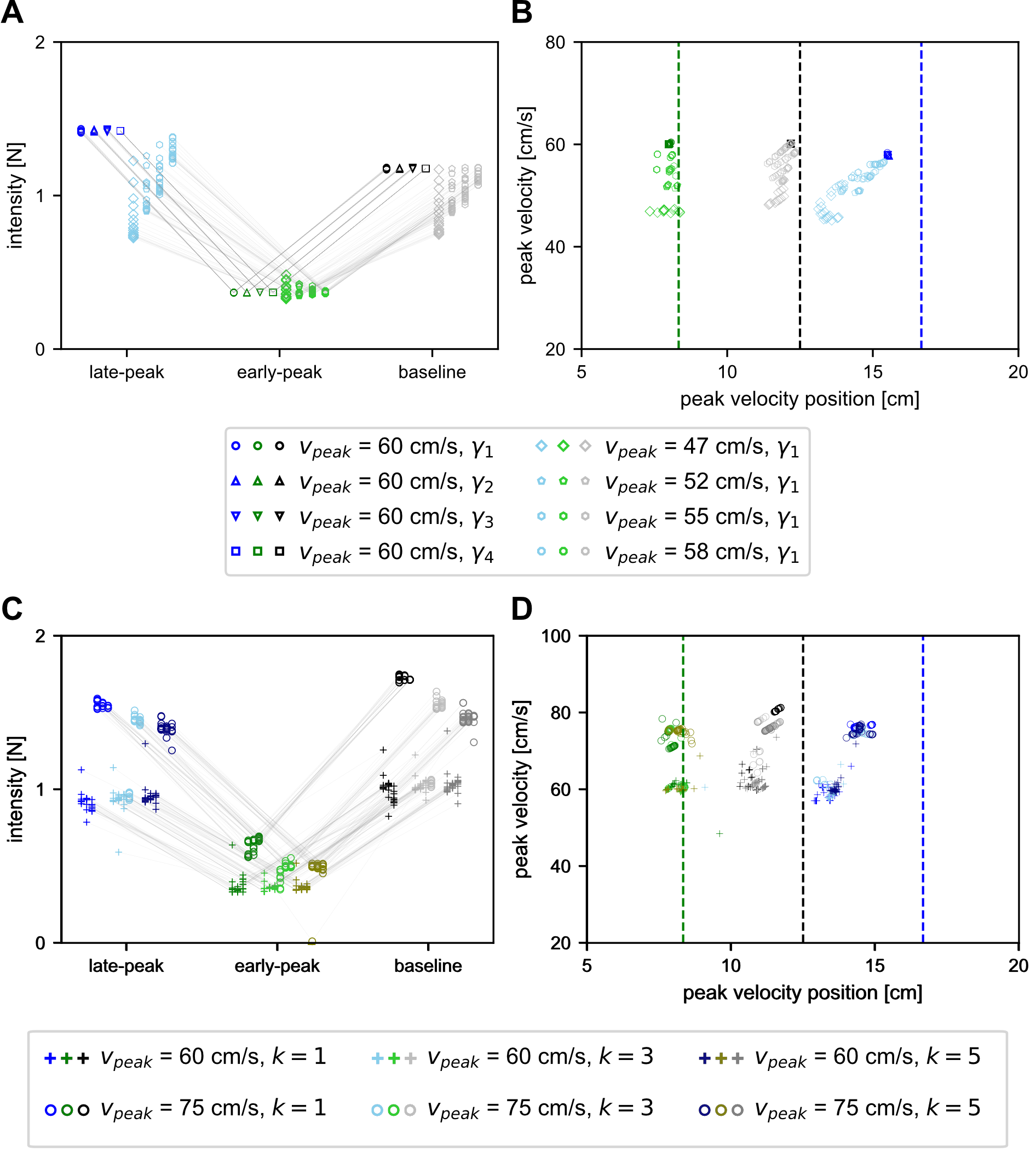}
	\caption{Cost sensitivity analysis for the separable and non-separable mixed-horizon OFC models. \textbf{A.} Simulated feedback intensity for different kinematic cost functions for the separable model. Every different cost function is represented by three points (blue, green and black), connected via a grey line. Each point represents a mean simulated feedback intensity for the respective simulated condition across five perturbation locations. Different marker styles represent a different family of kinematic cost functions. Overall the relative variation across different experimental conditions is consistent for a variety of cost functions. \textbf{B.} Simulated separable model kinematics for different kinematic cost functions represented as a combination of peak velocity location and magnitude. Altering the cost function affects the separation of the kinematic profiles where for some cost functions the baseline and the late-peak condition converge towards similar kinematics. Vertical dashed lines indicate desired location of peak velocity for each condition \textbf{C.} Simulated feedback intensities and \textbf{D.} simulated kinematics for the non-separable model. Different marker styles and colors represent different families of cost functions and noise levels respectively.}
	\label{fig:costsummary}
\end{figure}
\subsection{Cost sensitivity}

The optimisation sensitivity analysis presented above is aimed towards testing the reliability of our model.
Specifically, it demonstrates that given the actual parameters and cost functions that we use to obtain the final output, our model's behaviour can be reliably replicated, even though the parameter solution is numerically different.
However it is also important to test whether the behaviour of our model is a result of model structure and the algorithm itself, or if it is heavily influenced by the optimisation settings (i.e. a cost function).
Thus, in addition to optimisation sensitivity we also analyse the model sensitivity to different cost functions.
Particularly, as there is no variability in cost function choice for the infinite-horizon part of the model (as the desired output is straightforward), we will focus on the kinematic cost function relative weights $\gamma_1, \gamma_2, \gamma_3, \gamma_4$ (Equation \ref{met:eqn5}), and their effect on the model behaviour.

We performed this sensitivity analysis in the context of the model for our earlier study \cite{Cesonis2020}.
Here two alternatives are possible when we optimise the model parameters for different weights $\gamma_1, \gamma_2, \gamma_3, \gamma_4$.
On the one hand, modifying the kinematic cost function $\Gamma$ could impact the kinematics and dynamics of each condition independently, and as a result this would offset the relative regulation of feedback intensities across these conditions.
This would imply that the observed experimental behaviour replicated by our model is mostly influenced by the cost function used to generate a specific movement.
On the other hand, it is also possible that modifying the kinematic cost function will impact the kinematics and dynamics of the model in a way that the relative regulation of feedback intensities remains similar. 
This result would imply that the observed experimental behaviour is influenced by the underlying model structure, and not by the choice of the control parameters.

We performed the cost sensitivity analysis by first setting the kinematic cost function $\Gamma$ to its default value, with ($\gamma_1, \gamma_2, \gamma_3, \gamma_4$) = {(4, 4, 0.25, 25).}
In turn, we then varied each of the $\gamma$s within their boundary range while keeping the other three fixed.
We chose the boundary ranges for {$\gamma_1 \in [0.125, 16]$, $\gamma_2 \in [0.125, 16]$, $\gamma_3 \in [0.125, 16]$ and $\gamma_4 \in [0.25, 400]$.}
Furthermore, we also tested different values of $v_{peak, \, data}$ in the range {[47, 75] cm/s. }

For the separable model we initially tested {80} different cost functions {(20 for each $\gamma_{1-4}$, evenly spaced within the boundary range, with $v_{peak, \, data} = 60$ cm/s).}
For each simulation we looked at the kinematic features (location and magnitude of peak velocity) and dynamics (qualitative distribution of simulated feedback intensities, as well as mean intensity per condition).
The summary of these simulation results is shown in Figure \ref{fig:costsummary}A,B.
Generally, the results indicate that for separable models, within the range of parameters tested, varying parameters $\gamma_{2-4}$ has no effect on model behaviour, while changing $\gamma_1$ influences the peak velocity magnitudes slightly.
However, independent of the cost function, the model maintains the relative regulation of the three kinematic conditions producing consistent kinematics that satisfy the task requirements (Figure \ref{fig:costsummary}A).

As we only observed the variance in model behaviour within the modulation of $\gamma_1$, we further tested the effect of different target peak velocities in range [47, 60] cm/s and their relative weights (Figure \ref{fig:costsummary}AB).
By changing the target peak velocity magnitude within the cost function we alter the model behaviour in two ways.
First, reducing the peak velocity changes the movement kinematics. 
While the early-peak velocity condition can be successfully optimised to still produce the target kinematics, both baseline and late-peak conditions deviate further from the target kinematics (Figure \ref{fig:costsummary}B). 
Second, the movement dynamics (feedback intensities) also change.
The early-peak condition continues to produce weaker responses than the baseline condition, but the late-peak condition response decreases to the point where it is no longer upregulated in comparison to the baseline.
It is important to note, that although we altered the kinematic cost function, we did not alter the movement duration when changing this function -- the movement duration is solely produced by the infinite-horizon part of the model.
Thus, by changing the target velocity we demonstrate the model behaviour when moving away from its designed optimum point.

We repeated the sensitivity analysis for the non-separable model in the same way as for the separable model, except that we only simulated 10 equally spaced values for each $\gamma_{1-4}$ due to increased computational complexity (Figure \ref{fig:costsummary}CD).
In addition, we only analysed two select peak velocities: $v_{peak, \, data} = 60$ cm/s, and $v_{peak, \, data} = 75$ cm/s. 
The 60 cm/s were selected based on our results for the separable model providing the best separation among the kinematic profiles at this velocity (Figure \ref{fig:costsummary}B).
The 75 cm/s provides even better separation across the kinematics for the non-separable model (Figure \ref{fig:costsummary}D).
For both peak velocities we performed our simulations at the three different noise levels.
Importantly, here again we do not observe any meaningful differences across different noise levels in terms of the relative control gain regulation or kinematic targets (peak velocity position or magnitude, Figure \ref{fig:costsummary}C,D).
Overall our cost sensitivity analysis shows that, within the range of cost functions that we tested, the model behaviour remains consistent and thus the results are not strongly affected by the choice of the specific kinematic cost function.

\section{Results}
Our sensitivity analysis demonstrated that our models can produce kinematics and dynamics with some systematic differences that depend on the parameters and cost functions. 
The flexibility of our model allows the simulation of many different types of perturbations or movement conditions.
To demonstrate and test the generality of our mixed-horizon model, we can simulate specific conditions from previous studies to compare against experimental results. 
Here we use our model to replicate the experimental behaviour of three previous studies \cite{Cesonis2020, Dimitriou2013, Franklin2016}.
To do this, we now select a final set of parameters to model our data.
Specifically, we select these parameters to fit study 1, and then apply these same parameters to study 2 and study 3, unless the difference in behaviour requires a specific update (any differences detailed below).

\subsection{Study 1}

\subsubsection{Separable mixed-horizon OFC}

\paragraph{Final model parameters}
In our sensitivity analysis we already demonstrated that for the separable model there exists a set of values ($\omega_p$, $\omega_v$, $\omega_f$, $\omega_r$), with which our mixed-horizon OFC produces movements with durations matching our earlier experiment \cite{Cesonis2020}.
However, for each type of kinematics, in order to produce a required movement, three different parameter values are required ($\omega_p$, $\omega_v$ and $\omega_r$, as $\omega_f$ = $\omega_v$*10).
From the human motor control perspective this becomes challenging -- optimising for, and learning a new set of control parameters for each new movement would be computationally expensive.
On the other hand, it could be faster to produce a novel movement if a single hyper-parameter could be used to switch between these different kinematics.

\begin{figure}[t!]
	\centering
	\includegraphics[width=\linewidth]{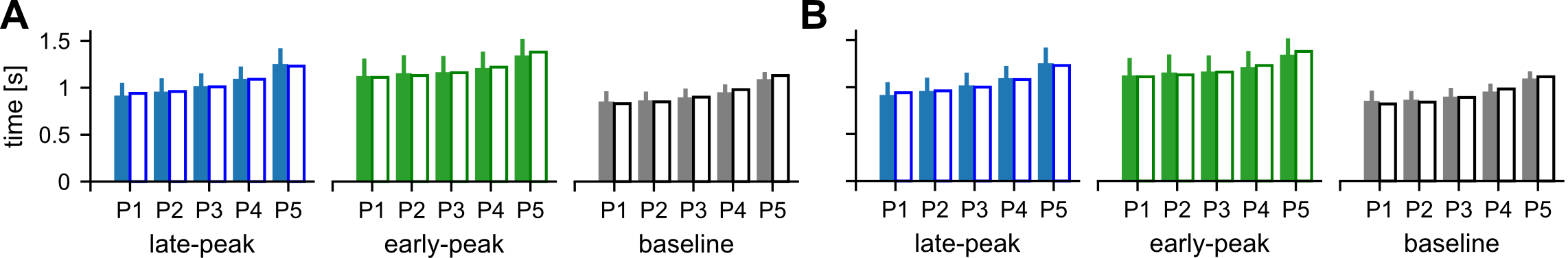}
	\caption{Movement duration comparison between the data (solid bars) and the final mixed-horizon model (white bars). Error bars on the data show 95\% CI. No error bars are shown for the model results, as the durations result from a single simulation with final model parameters. P1-P5 indicate movements with different perturbation onsets.  \textbf{A.} Comparison for the separable model. \textbf{B.} Comparison for the non-separable model.}
	\label{fig:st1_1}
\end{figure}

Figure \ref{fig:sep_opt_sen} demonstrates the relationship of the optimum parameter spaces across all three different kinematics for the infinite-horizon part of the model.
As parameters for each condition are distributed in a plane, such that its projection to the $\omega_p$-$\omega_v$ plane is a line, we can select parameters $\omega_p$ and $\omega_r$, common across all three conditions, and find a value of $\omega_v$, unique for each condition.
Indeed, by arbitrarily selecting ($\omega_p$, $\omega_r$) = (1500, 0.006) and performing optimisation on $\omega_v$ for each condition we found $\omega_v$ = 0.106, $\omega_v$ = 0.0713 and $\omega_v$ = 0.0501 for the early peak, late peak and baseline kinematics respectively.
With these control parameters, our model produced movement durations that fit within the 95\% confidence intervals of the experimental durations, showing that infinite-horizon OFC could be used to estimate the duration of the movement for the finite-horizon control (Figure \ref{fig:st1_1}A).

In order to now select final model parameters for the finite-horizon part of the model, we first need to select the values for the kinematic cost function $\Gamma$ (Equation \ref{met:eqn5}).
Figure \ref{fig:costsummary} represents the summary of resultant kinematics and relative regulation of dynamics for each $\Gamma$ tested.
In order to best represent the experimental design and requirements for participants, we selected a kinematic function that maximises the separation of the three velocity profiles (i.e. the peak velocity locations are closest to their experimental requirements).
As a result, the best parameters of the kinematic function suitable for simulation of the experimental data are ($\gamma_1, \gamma_2, \gamma_3, \gamma_4$) = (4, 4, 0.25, 25), with the $v_{peak, \, data} = 60$ cm/s.

Unlike the infinite-horizon part, the finite-horizon part of the model does not require a different set of parameters to produce each of the three different kinematics.
Instead, the early-peak and late-peak conditions are generated with the same base parameters $\omega_p$, $\omega_v$, and $\omega_f$. The $\omega_r$ is modulated according to Equation \ref{met:eqn8} to produce the skew in the velocity profile over time, where its mean over the movement duration is identical across the three conditions.
In terms of the optimisation to find the parameters for the two conditions, we fixed $\omega_p$, $\omega_v$, $\omega_f$ and $\omega_r$ to their baseline values and found the $p$, $q$ and $r$ (Equation \ref{met:eqn7}) that produced closest to desired kinematics.
{As a result, for the baseline condition we used ($\omega_p$, $\omega_v$, $\omega_f$, $\omega_r$) = (329.5, 9.859, 98.59, 5.568 $\times$ 10$^{-6}$).
For early peak condition, in addition to the baseline parameters we used ($p$, $q$, $r$) = (52.72,  50.37, 741.2), and for late peak condition we used ($p$, $q$, $r$) = (-13.24,  21.08, 347.5).}

\begin{figure}[p!]
	\centering
	\includegraphics[width=\linewidth]{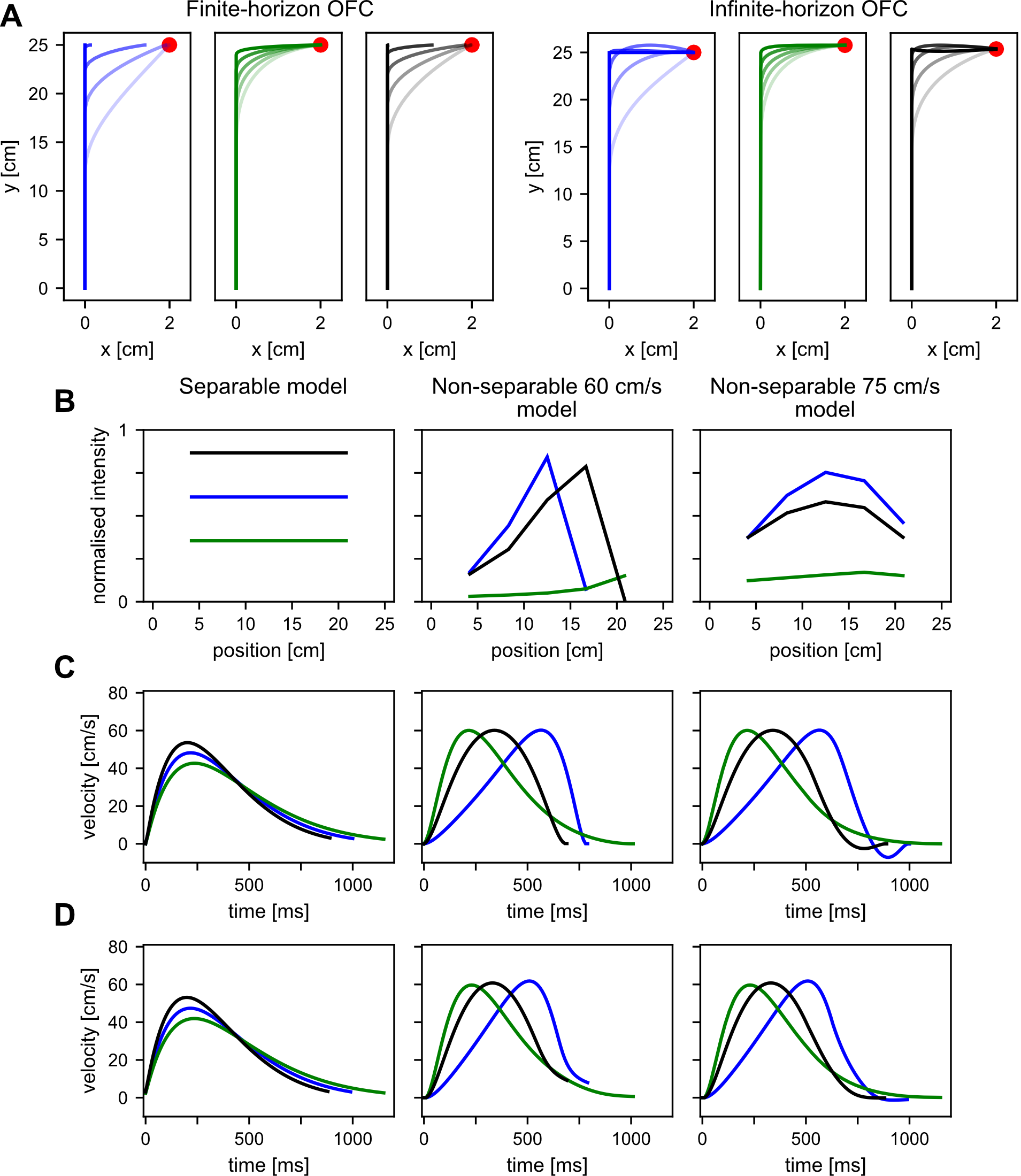}
	\caption{Comparison of simulation results across different models. Blue, green and black lines represent late-peak, early-peak and baseline conditions respectively. \textbf{A.} Kinematic trajectory comparison between finite-horizon model (left) and an equivalent mixed-horizon model (right). Mixed-horizon model always converges to the target while the finite-horizon model produces undershoots due to the lack of time to complete the movement after the perturbation. Differently shaded lines indicate the five different perturbation onset locations. Three panels in each figure indicate early-peak, late-peak and baseline conditions. \textbf{B.} Temporal evolution of feedback intensities for infinite-horizon (left), finite-horizon (middle) and mixed-horizon (right) separable models across different kinematic conditions. Human-like temporal evolution is only observed in the mixed-horizon model. \textbf{C.} Velocity profiles in non-perturbed movements across three different kinematic conditions for the three separable models. The separation of different kinematics is not achieved by the infinite-horizon model. \textbf{D.} Velocity profiles during non-perturbed movements for non-separable models. Importantly, compared to the separable models, the velocity profiles of non-separable models are skewed towards the beginning of the movement, producing lower separation among different kinematic conditions.}
	\label{fig:ofccomp}
\end{figure}

\paragraph{Model behaviour}
Here we propose the mixed-horizon OFC implementation as an extension of the finite-horizon OFC and infinite-horizon OFC when simulating perturbed movements.
The goal of this mixed-horizon implementation is to address the limitations that the fixed horizon OFC encounters when modelling such movements.
First, we compared the movement trajectories between the finite-horizon OFC controlled movement and the mixed-horizon OFC controlled movement (Figure \ref{fig:ofccomp}A).
We simulated movements where the target was perturbed laterally at one of five different distances along the movement, requiring a correction to reach the target.
For the finite-horizon OFC generated trajectories we observe the stereotypical undershooting for later perturbations, arising from the fact that the perturbation occurs near the end of the planned movement. 
As a result, it is more optimal for the controller to pay an end-point penalty, than to produce a vigorous and effortful movement in the short remaining time.
On the other hand, the mixed-horizon controller produces movements that always converge to the target, consistent with the task requirements and human behaviour.   
Note that the infinite-horizon controlled movements also always converge to the target due to the nature of the controller and thus are not shown in Figure \ref{fig:ofccomp}A.

In terms of the model dynamics, the mixed-horizon OFC model qualitatively replicated the feedback intensity profiles of human participants. 
In the original study, human participants regulated the intensity of visuomotor feedback responses to lateral cursor jumps, both within the same movement, as well as across different movement kinematics.
Our mixed-horizon OFC model shows similar behaviour.
When compared to the infinite or finite-horizon models, our mixed-horizon model shows a visible improvement (Figure \ref{fig:ofccomp}B).
In addition, while the mixed-horizon model shows similar qualitative behaviour as the time-to-target model (Figure 7C in \cite{Cesonis2020}), it requires fewer inputs and is thus more useful when modelling novel behaviours.

Finally, we compared the separable mixed-horizon model to infinite-horizon and finite-horizon models in their ability to modulate the velocity across three different experimental conditions (Figure \ref{fig:ofccomp}C).
In terms of peak velocity separations, both the finite-horizon and mixed-horizon models produced the required modulations in the shape of the velocity profiles. 
However, the infinite-horizon OFC could not produce such modulation.
Instead, all three conditions in the infinite-horizon OFC manifested in similar velocity profiles with only differences in peak velocity to produce different movement durations.
In addition, the velocity profiles were always positively-skewed.
However, even with such skewed profiles of the infinite-horizon OFC as a component, the mixed-horizon model is able to capture the appropriate movement durations, condition dependent velocity profiles, and human-like modulation of feedback gains.

\subsubsection{Non-separable mixed-horizon OFC}

The non-separable mixed-horizon OFC model differs from the previously presented separable model in its implementation.
The main difference between the two models is that the non-separable model contains the multiplicative noise factor, which is equivalent to control dependent noise in humans.
Thus, the control policies that emerge from this model may be different than the ones used when no such noise is present.
However, at the operational level both types of models are used in a similar fashion, and hence similar behaviours can be simulated. 
As a result, here we replicate the same experimentally tested human behaviours as we did with the separable model.
Importantly, as we already described the effect of different control-dependent noise levels on the model behaviours, we will only use the control dependent noise with a scaling factor $k=1$ for further simulations.

\paragraph{Final model parameters}

The final parameters  for the study 1 model were selected based on our previous sensitivity analysis, using the same criteria as for the separable model.
First, based on results shown in Figure \ref{fig:nonsep_opt}A we know that for each kinematic condition the optimum parameters are distributed in a plane, similarly to the separable model parameters.
Thus, for each kinematic condition we fixed two out of the three controller costs {($\omega_p$, $\omega_r$) = (2000, 0.009) and optimised for the $\omega_v$ for each condition so that the resultant movement durations match those in the experimental data.
For late-peak, early peak and baseline kinematic conditions we obtained $\omega_v$ = 0.0969, $\omega_v$ = 0.144 and $\omega_v$ = 0.0663 respectively.}
With the resulting controller parameters our model produced movement durations that fit within the 95\% confidence intervals of these durations in experimental data, showing that infinite-horizon OFC could be used to estimate the duration of the movement for the finite-horizon control (Figure \ref{fig:st1_1}B).

In order to determine the controller costs for the finite-horizon part of the model we followed the same strategy as for the separable model.
Particularly, we first selected a kinematic cost function $\Gamma$, and then used it to optimise for the best fit values for controller parameters.
As our sensitivity analysis showed that no significant changes in mean behaviour of the model are introduced with the different levels of control-dependent noise, we only analysed the conditions where $k=1$.
However, with the overall presence of control dependent noise we no longer have a clear-cut best $\Gamma$ to describe our kinematics.
Instead, we observe a trade-off between peak velocity location and magnitude for the late-peak velocity condition {(Figure \ref{fig:costsummary}D)}.
Hence, to fully analyse the model behaviour we will look at the best-fit kinematic cost function for $v_{peak} = 60$ cm/s and $v_{peak} = 75$ cm/s, as the former sufficiently meets the peak velocity criterion while the latter produces the best separation between the peak locations.

First, we chose the values for the kinematic cost function $\Gamma$ as ($\gamma_1, \gamma_2, \gamma_3, \gamma_4$) = (4, 4, 0.25, 25) and the $v_{peak, \ desired} = 60$ cm/s.
{With these values we then found the baseline parameters ($\omega_p$, $\omega_v$, $\omega_f$, $\omega_r$) = (1307, 10.7, 107, 2.27$\times 10^{-4}$).
In addition to the baseline parameters, for the early peak condition we used ($p$, $q$, $r$) = (52.66, 81.05, 772.6), and for the late peak condition we used ($p$, $q$, $r$) = (-13.45, 60.29, 322.3).
The best-fit control parameters for $v_{peak, \ desired} = 75$ cm/s were obtained by optimising the kinematic cost function $\Gamma$ with ($\gamma_1, \gamma_2, \gamma_3, \gamma_4$) = (1.44, 4, 0.25, 25), which yielded ($\omega_p$, $\omega_v$, $\omega_f$, $\omega_r$) = (627.6, 2.243, 22.43, 2.321$\times 10^{-6}$) for the baseline condition, and ($p$, $q$, $r$)$_{early}$ = (32.24, 18.61, 520.4) and ($p$, $q$, $r$)$_{late}$ = (-14.83, 37.55, 260.5).}

\paragraph{Non-separable model behaviour}
In order to evaluate the model behaviour for the non-separable model, we compared the kinematics and feedback intensities of this model with the separable model, and with the experimental feedback intensities \cite{Cesonis2020}.
First, the velocity profiles produced by non-separable infinite-horizon OFC still showed similar lack of variation as in the separable case (Figure \ref{fig:ofccomp}D).
In addition, the finite-horizon and mixed-horizon models produced velocity profiles that were positively-skewed compared to the separable models, demonstrating the effect of the control dependent noise.
That is, with control-dependent noise and a peak velocity target, the controller chooses to reach this velocity earlier in the movement to be able to have lower velocity and lower noise near the target.
While qualitatively this change appears minor (Figure \ref{fig:ofccomp}CD), quantitatively this results in earlier perturbation onset times (as our model moves faster in the beginning of the movement) and subsequently longer times-to-target for each perturbation.
In turn, longer times-to-target for these perturbed movements should result in weaker feedback response intensities, particularly for baseline and late-peak velocity conditions where the effect of such velocity front-loading is the largest.

\begin{figure}[t!]
	\centering
	\includegraphics[width=0.82\linewidth]{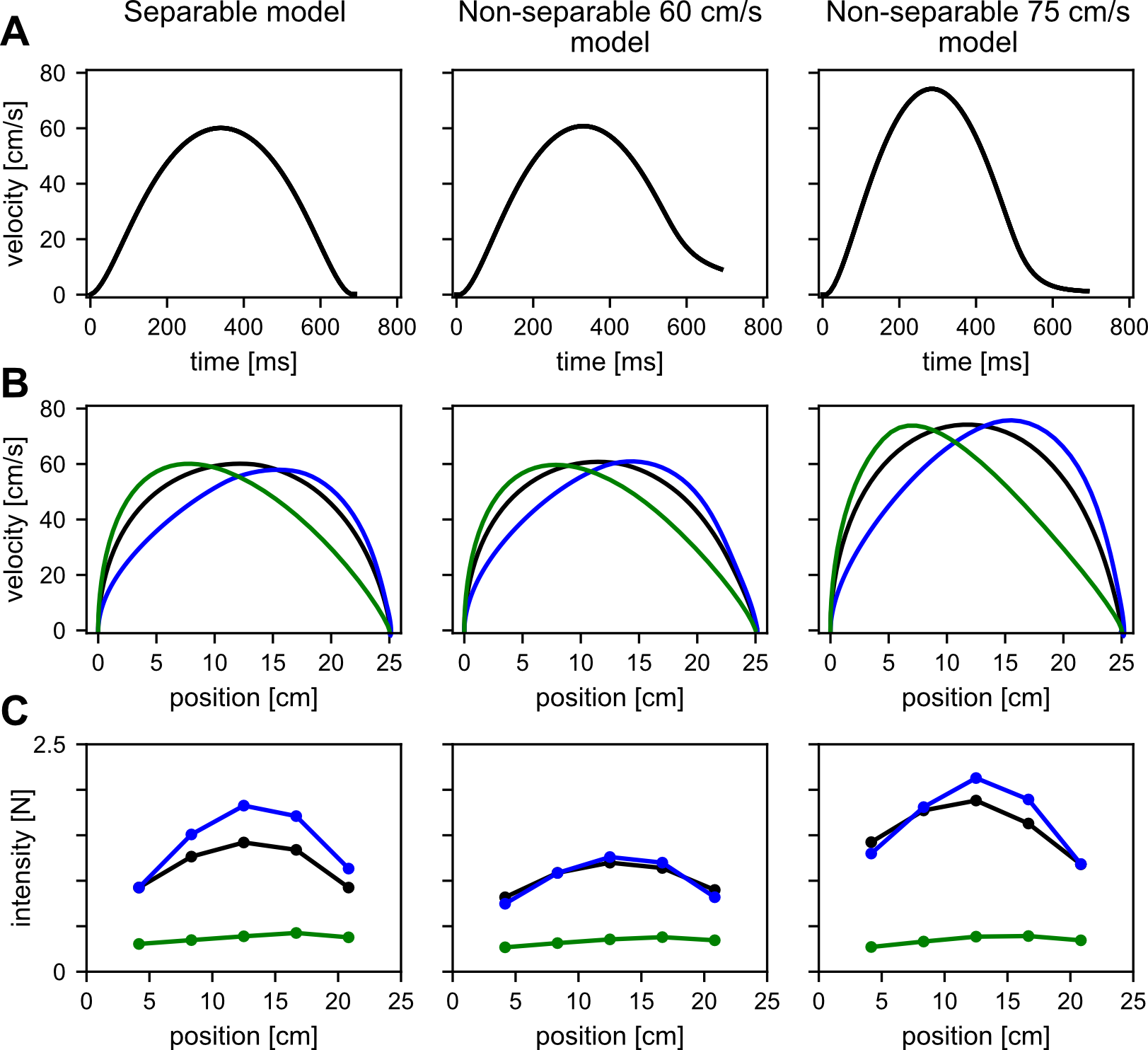}
	\caption{Comparison of separable and non-separable model simulations of kinematics and feedback intensities for study 1. Left, middle and right columns demonstrate the results for the separable, non-separable 60 cm/s velocity target, and non-separable 75 cm/s velocity target models. \textbf{A.} Forward velocity as a function of time for the finite-horizon part of the mixed-horizon model. The separable model produces a movement that successfully fulfils requirements of both peak velocity and zero final velocity. In the non-separable models, the noise introduces a trade-off between either finishing the movement with non-zero velocity (60 cm/s model) or increasing the peak velocity (75 cm/s model). \textbf{B.} Position-velocity dependency. All three models successfully produce the required velocity profiles across the conditions. Note that the 75 cm/s model produces a clearer separation of peak velocities across the conditions compared to the non-separable 60 cm/s model. \textbf{C.} Simulated feedback intensities. The separable model successfully replicates experimental results. Non-separable models show limitations in successfully up-regulating the late-peak velocity condition, but capture the down-regulation of the early-peak velocity condition.}
	\label{fig:St1_2}
\end{figure}

Second, we compared just the baseline kinematics of the finite-horizon parts of the separable and non-separable models, as this demonstrates the default, non-perturbed behaviour (Figure \ref{fig:St1_2}A).
Here the separable model produces the classic bell-shaped velocity profile peaking at the required velocity of 60 cm/s and stopping at the target.
In contrast, the 60 cm/s peak velocity requirement for a non-separable model produces a movement that satisfies the required peak velocity, but fails to stop at the target.
An attempt to reduce this terminal velocity by increasing $\gamma_3$ would only increase the priority of stopping by compromising some of the other requirements -- either by shifting the peak velocity (magnitude or location), or by overshooting or undershooting the target.
This is illustrated by the kinematics of the 75 cm/s condition, where an increase in the peak velocity allowed the model to stop at the target.
The addition of control dependent noise, therefore, changes the default kinematics of the model, particularly affecting the velocity profiles which in turn can affect the dynamics of the system.

Finally, we compared the overall kinematics and feedback intensities across separable, non-separable 60 cm/s and non-separable 75 cm/s models (Figure \ref{fig:St1_2}BC).
Importantly, all three models were able to reproduce the different kinematic requirements for each condition.
In addition, in all three models the response intensities for the early-peak condition were downregulated compared to the other two conditions.
However, as expected from the control-dependent noise effects on kinematics, the major difference between the two non-separable models compared with the separable model and experimental data was in the relative regulation of baseline and late-peak response intensities.
While the experimental data and separable model showed a clear upregulation for the late-peak intensities, this difference is reduced significantly for the non-separable models.
Specifically, for the 60 cm/s model the two profiles are virtually indistinguishable, while for the 75 cm/s model this difference is slight but consistent.
Notably, it appears that this absence of regulation in the non-separable 60 cm/s model results from the late-peak condition, as the baseline feedback responses are just slightly lower than for the separable model (Figure \ref{fig:St1_2}C).  

Overall, the presence of control dependent noise in the mixed-horizon model distorts the experimentally observed regulation of feedback intensities across the three kinematic conditions (Figure \ref{fig:St1_2}C, middle), which was successfully simulated without the control dependent noise (Figure \ref{fig:St1_2}C, left).
The source of this mismatch is due to the regulation of the late-peak velocity condition, which required unusual and unnatural movements and is thus unlikely to be encountered in typical model applications.
Finally, for this and the other conditions, our mixed-horizon model produced the variation in temporal evolution closer to that of the human participants than other model candidates (Figures \ref{fig:ofccomp}B and \ref{fig:St1_2}C).

\subsection{Study 2}

\subsubsection{Separable mixed-horizon OFC}
\paragraph{Final model parameters}

Here we modelled the experimental paradigm of \cite{Dimitriou2013}, experiments 2 and 3, where the reaching target was perturbed in the forward movement direction simultaneously with, or 100 ms before, the cursor perturbations used to measure the feedback intensities.
Specifically, we modelled 6 conditions: 2 $\times$ cursor perturbations, 2 $\times$ target perturbations (near to far and far to near), and 2 $\times$ combined target and cursor perturbations.
As all conditions appeared in the same experiment in a randomised order, we used the same control parameters across all conditions.

The infinite-horizon part of the model is used initially by the controller in order to produce movements of an appropriate duration.
Thus, we optimised the infinite-horizon control parameters for modelling this study by fitting the modelled movement durations to experimentally recorded movement durations for each experiment.
{For experiment 2 the obtained values were ($\omega_p$, $\omega_v$, $\omega_f$, $\omega_r$) = (5448, 0.05285, 0.5285, 0.002460).
For experiment 3 the values were ($\omega_p$, $\omega_v$, $\omega_f$, $\omega_r$) = (5606, 0.04114, 0.4114, 0.0003922).}
However, for the finite controller we used the identical control parameters to study 1 as the task requirements such as movement distance, speed and perturbation size were comparable.

\begin{figure}[t!]
	\centering
	\includegraphics[width=0.82\linewidth]{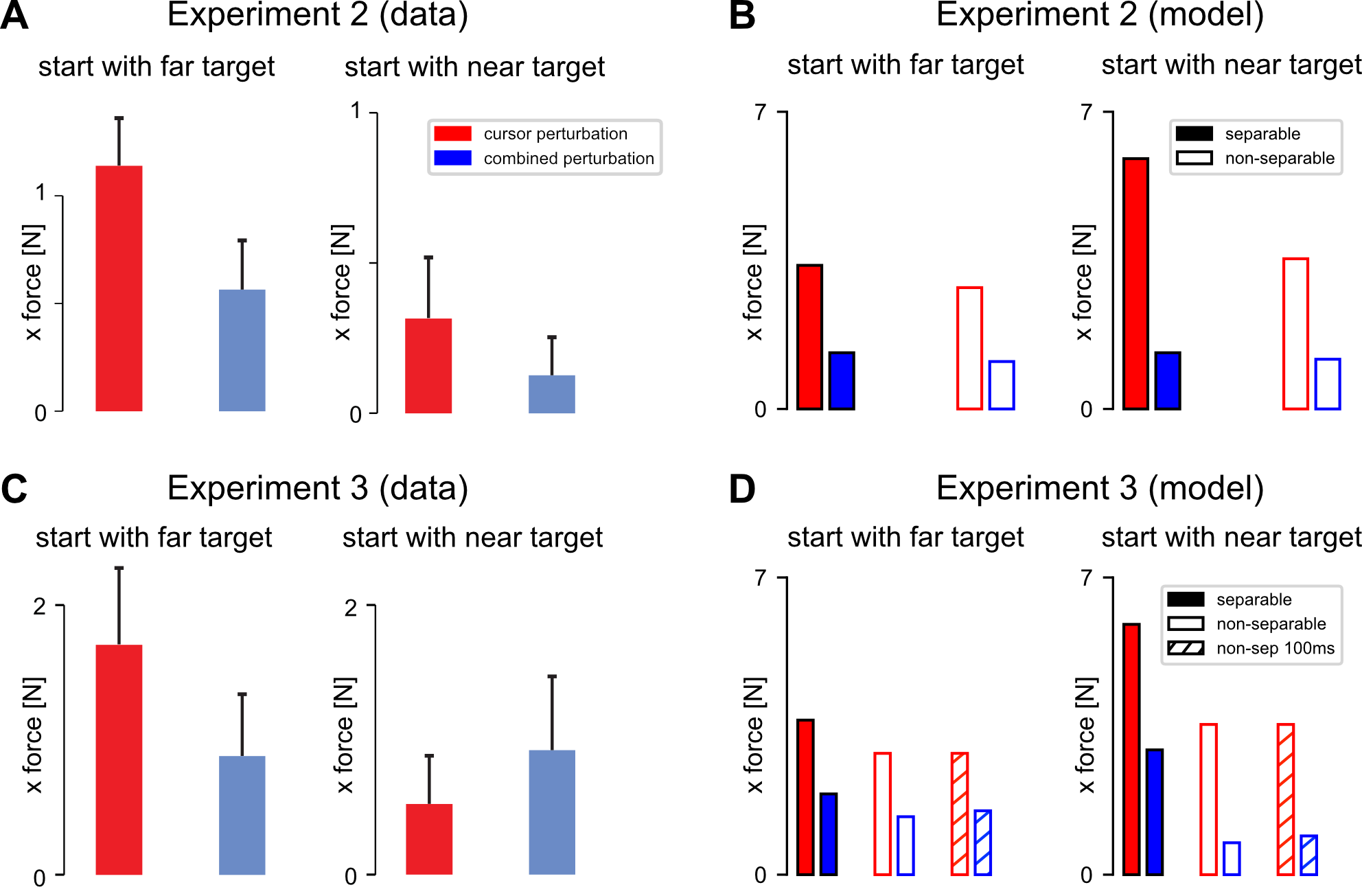}
	
	\caption{Data and model simulation comparisons for study 2. \textbf{A,C.} Figures adapted from the original published paper \cite{Dimitriou2013} showing the feedback response intensities in experiment 2 (simultaneous perturbations) and 3 (cursor and target perturbations separated by 100 ms) respectively. Red bars show responses to cursor-only perturbations while blue bars show responses to combined perturbations. \textbf{B, D.} Respective model simulations with separable and non-separable models. For experiment 3 we also include another non-separable model where the perturbation onset times are adjusted to time-match the perturbations in the data.}
	\label{fig:dim_sep}
\end{figure}

\paragraph{Model behaviour}
We compared the simulated feedback responses of only cursor perturbations to the combined cursor and target perturbations.
In experiment 2 of the original study \cite{Dimitriou2013}, cursor and target perturbations occurred simultaneously.
Participants produced stronger corrective responses to perturbations of only the cursor, than when both the target and the cursor were perturbed (Figure \ref{fig:dim_sep}A, Figure 3FG in \cite{Dimitriou2013}).
Our mixed-horizon model similarly produced weaker corrective responses for simultaneous cursor and target perturbations, compared to just cursor perturbations (Figure \ref{fig:dim_sep}B).
However, while experimental data showed overall stronger responses for conditions where the starting target was "far" compared to when the starting target was "near", our model simulations showed the opposite.

In experiment 3 of the original study, target jumps (when present) were induced earlier in the movement than the cursor jumps.
For the conditions with the "far" starting target, the results of this experiment were similar to the results of the equivalent conditions from experiment 2 (Figure \ref{fig:dim_sep}C, Figure 4FG in \cite{Dimitriou2013}).
Our model also produced such responses, with corrections to cursor perturbations being stronger than the combined cursor and target perturbations (Figure \ref{fig:dim_sep}D).
However, participants produced stronger responses to the combined perturbations than only to the cursor perturbations when the starting target was "near", which was not consistent with the results of our model.

\subsubsection{Non-separable mixed-horizon OFC}

\paragraph{Final model parameters}
As previously, in the separable model, we fit the infinite-horizon part of the model to reproduce the movement durations recorded by experimental participants.
For both experiment 2 and experiment 3 we set {($\omega_p$, $\omega_v$, $\omega_f$, $\omega_r$) = (3585.9, 0.04073, 0.4073, 0.009247) for the infinite models.}
For the finite-horizon part of the model we maintained the parameters we previously found for study 1 60 cm/s condition. 
The 60 cm/s parameters were chosen over the 75 cm/s parameters because for both settings the baseline condition was producing kinematics consistent with the task requirements -- only the late-peak condition was causing discrepancies between models and data.
As for both study 2 and study 3 we only use the natural, baseline kinematics, here we opted for the 60 cm/s as it is closer to the kinematics in the data.

\paragraph{Model behaviour}

We implemented the non-separable model to simulate experiment 2 and experiment 3 of study 2, similar to the separable model simulations.
Qualitatively the non-separable model produced similar results as the separable model (Figure \ref{fig:dim_sep}BD).
Specifically, conditions with combined target and cursor perturbations produced weaker responses to the cursor perturbations, and conditions that started with the near target also produced stronger responses to isolated cursor perturbations.
In addition, the non-separable model, similar to the separable model, also failed to replicate the response modulation when the cursor and target were perturbed at different times (starting with near target, experiment 3).
Instead, similar to the separable model, the combined perturbation produced weaker responses than the isolated cursor perturbation.

The original study describes combined perturbations in experiment 3 as "separated by ~100 ms", however both of these perturbations are induced via the hand position crossing an onset location.
Thus, such time dependency is dependent strongly on the shape of the velocity profile.
Indeed, in our simulations these perturbations were only separated by 60-70 ms which could have influenced the results.
We performed another simulation where instead of perturbing the target once the cursor crosses 10.5 cm distance from the start, we induced this perturbation at 8.5 cm resulting in 100 ms delay between the target and cursor perturbations.
Still this modification did not change our results significantly, with combined perturbation still producing lower gains than isolated perturbation for the near target condition (Figure \ref{fig:dim_sep}D).

\subsection{Study 3}
\subsubsection{Separable mixed-horizon OFC}
\paragraph{Final model parameters}
In order to simulate the control behaviour when subjected to different perturbation sizes we modelled a part of the experimental paradigm of \cite{Franklin2016}.
Particularly, we simulated the conditions where either a cursor, or a target, was perturbed by 1, 2 or 3 cm, at the mid-point in the forward movement.
In order for our modelled movement durations to match the experimentally recorded movement durations we again optimised the infinite-horizon part of the model, and we used the same set of infinite-horizon parameters across all conditions.
The obtained values were {($\omega_p$, $\omega_v$, $\omega_f$, $\omega_r$) = (2292, 0.05868, 0.5868, 0.006573).}
However, in terms of control requirements, movements in this study were similar to the ones in study 1 baseline condition, thus we used the previous parameters for the finite-horizon part of the model.

\begin{figure}[p]
	\centering
	\includegraphics[width=0.80\linewidth]{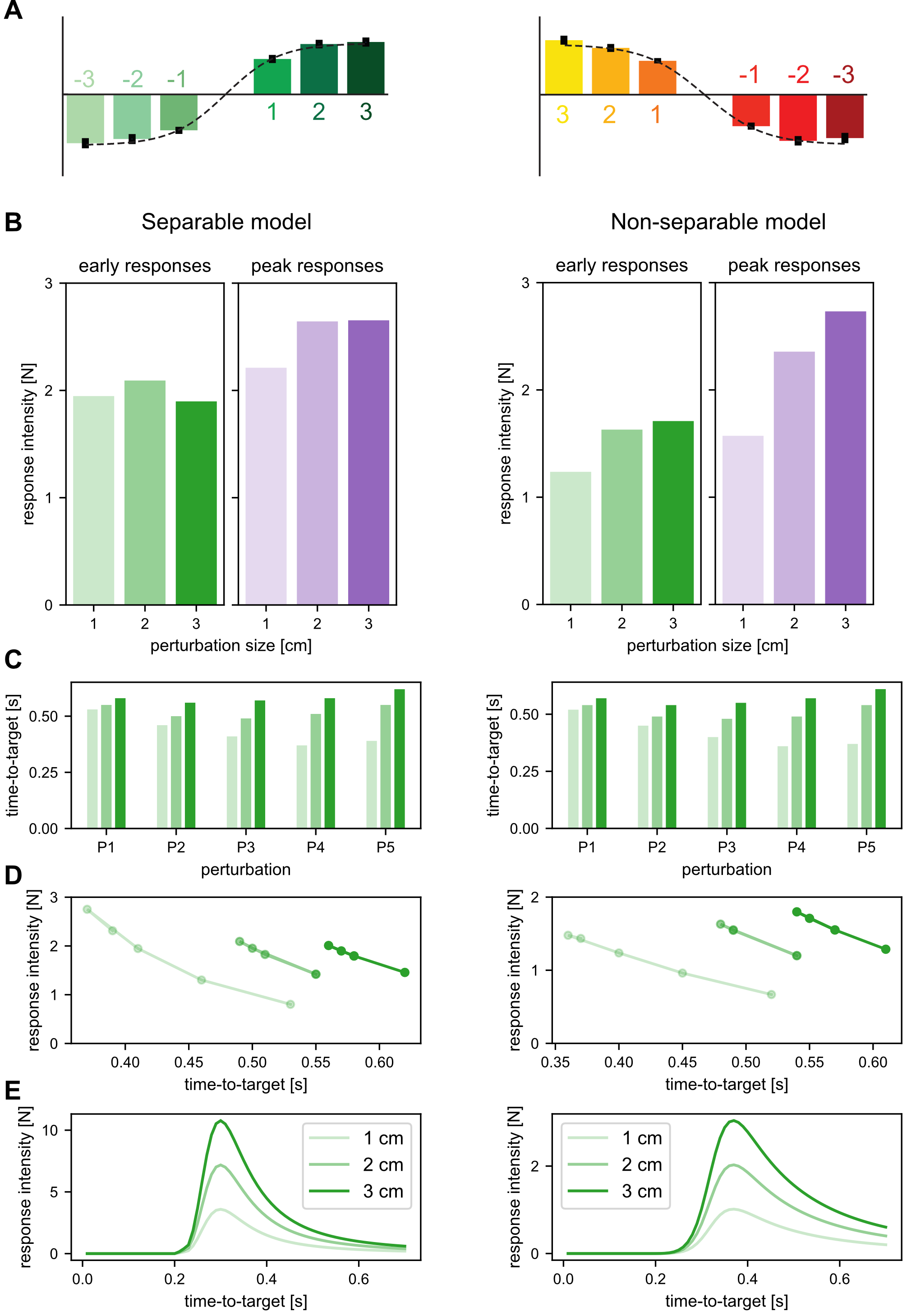}
	\caption{Data and model simulations for study 3. \textbf{A.} Experimental results. Early feedback intensity to a perturbation at 12.5 cm along the movement in relation to perturbation size in cm. Left plot shows the modulation to target perturbations while the right plot shows for the equivalent cursor perturbations. A non-linear saturation in response intensity is observed. Figure adapted from the original study \cite{Franklin2016}. \textbf{B-E.} Model simulations. Left half shows results for the separable model, right half for the non-separable model. \textbf{B.} Simulated intensities for different perturbation sizes. Green bars show early responses, purple bars show peak responses. Similar behaviour to experimental results is achieved by the model simulations. \textbf{C.} Simulated time-to-target for perturbations of sizes 1 cm, 2 cm and 3 cm, and with onset at one of the five locations P1-P5. Perturbation onset location and magnitude has a non-monotonic effect on the time-to-target. \textbf{D.} Same perturbation responses visualised against time-to-target. Perturbations with matching onset locations produce different times-to-target. \textbf{E.} Time-to-target effect on response intensity generalised to all perturbations simulated by the mixed-horizon model. Modelling results show a linear increase in response intensity to perturbations happening at the same time-to-target.}
	\label{fig:dfseparable}
\end{figure}

\paragraph{Model behaviour}
{In the original study, 1, 2, and 3 cm lateral perturbations in the middle of the forward movement resulted in visuomotor feedback intensities (early responses 170-230 ms) that did not scale linearly, but saturated or even reduced for larger perturbation sizes (Figure \ref{fig:dfseparable}A, Figure 1DE in \cite{Franklin2016}).
We simulated these conditions with our mixed-horizon model.
First, our feedback response intensities showed similar regulation -- responses to increasing perturbation sizes initially increased and then slightly reduced.
In the experiment, the late responses (370-430ms) corresponded roughly to the peak forces within the correction.
Therefore, we used the simulated peak response intensities as a proxy for these late responses. 
While these responses were stronger in intensity compared to the early simulated responses, they also saturated with increasing perturbation size, similar to the early human responses (Fig. \ref{fig:dfseparable}B).}

{
It is worth noting that, due to the linearity of our controller, one could reasonably expect the feedback intensities to scale linearly with perturbation size, particularly as the perturbations in the original experiment happen at the same time, speed and forward distance.
Therefore, it is important to understand how our linear, mixed-horizon OFC can produce non-linearly increasing responses with linearly increasing stimuli.
To illustrate the source of this behaviour, we can use the strength of the mixed-horizon implementation and simulate additional conditions (i.e. perturbations at different onsets) that provide broader context to the available data, but were not experimentally tested.
Thus, for each of three perturbation sizes (1, 2 and 3 cm) we additionally simulated four other perturbed movements: 1/6, 1/3, 2/3 and 5/6 of the distance along the movement, similar to study 1.
In total, at each of the five perturbation onset locations we simulated 3 perturbations, one of each magnitude (1 cm, 2 cm and 3 cm).
As the velocity requirements were identical for non-perturbed movements independent of the upcoming perturbation size, the perturbations at the same onset location therefore happened at the same position, same movement time, and same movement velocity.

Even though the movement kinematics and dynamics, as well as perturbation onset completely matched before the perturbations, introduction of the perturbation resulted in the remaining movement time (time-to-target) being differently adjusted by the controller after the perturbation.
Importantly, the resultant times-to-target were always longer for larger perturbations, which is an intuitive result, given that larger corrections were necessary to reach the target (Figure \ref{fig:dfseparable}C).
Also notably, perturbations of a specific magnitude occurring close to the target could extend the movement duration beyond the same perturbation occurring earlier in the movement.
For example, in some cases (e.g. perturbations at 2/3 and 5/6 of the movement) a perturbation that occurred closer to the target produced a longer time-to-target than an earlier perturbation with the same magnitude.
While this result is less intuitive, it is consistent with our previous experimental results (\cite{Cesonis2020}).

The time-to-target is an important parameter in the LQG and LQR implementation, as the control gains are computed in the backwards-pass calculation, where at every future time point, starting from the end of the movement, the movement cost-to-go and thus the appropriate optimal control gains are computed \cite{Todorov2005}.
Thus, we looked at our simulated feedback intensities for the newly induced perturbations with respect to the time-to-target (Figure \ref{fig:dfseparable}D).
Two important results were observed: first, there is a systematic relation between the time-to-target and the feedback intensities, with higher intensities produced for shorter times-to-target within these responses.
Second, due to the extension in movement time post-perturbation, the perturbations happen at different times-to-target, even when they occur at the same location, or same movement time.
Due to this difference we also observe a non-monotonic response regulation with perturbation onset location or onset time.
Thus, even though, due to the linearity of the controller, the responses that happen at the same time-to-target scale linearly with perturbation size as expected (Figure \ref{fig:dfseparable}E), the overall regulation of feedback intensities induced at the same location or onset time is non-linear (Figure \ref{fig:dfseparable}AB).
}

\subsubsection{Non-separable mixed-horizon OFC}
\paragraph{Final model parameters}

For the non-separable model of study 3 we again re-fit the infinite-horizon parameters so that the movement durations generated by our model match those in the data.
This resulted in the following infinite-horizon parameters: {($\omega_p$, $\omega_v$, $\omega_f$, $\omega_r$) = (2321, 0.05999, 0.5999, 0.001706).}
For the finite-horizon part of the model we maintained the parameters we previously found for study 1 60 cm/s condition and already used for study 2.

\paragraph{Model behaviour}
We did not observe any significant qualitative or quantitative changes in our simulations for study 3 when comparing the non-separable model results with previously simulated separable model results (Figure \ref{fig:dfseparable}B-E).
Thus, introducing control-dependent noise to this particular model did not have a behavioural effect, demonstrating consistency in the simulated behaviour.

\section{Discussion}
Here we built and analysed a novel implementation of the optimal feedback controller (OFC) for reaching movements -- a mixed-horizon OFC.
In order to do this, we not only evaluated the behaviour of the controller when applied to motor control problems, but also analysed the parameters that generate the optimal behaviour.
Specifically, our mixed-horizon OFC could successfully replicate kinematics as well as many dynamic features shown by human participants \cite{Cesonis2020, Dimitriou2013, Franklin2016}, that were not previously possible with either a finite-horizon or infinite-horizon OFC.
In addition, our model allowed us to test new hypotheses and alternative explanations to previously published results that could further strengthen original conclusions (study 2) or find new explanations to previously observed interesting features (study 3).
Furthermore, by testing the model sensitivity we demonstrated the distinction between fundamental behavioural differences across experimental conditions of study 1, and showed that these differences are not the effect of deliberate model fitting, but maintained across different cost functions. 

Prior implementations of either a finite-horizon OFC or infinite-horizon OFC had strong limitations in predicting human behaviour.     
First, a conventional finite-horizon OFC requires a movement duration as an input.
Real world movements on the other hand are almost always free from a specified movement duration.
For example, no-one reaches for their coffee cup in the morning with a specified duration of 550 ms. 
Even if we consider laboratory experiments, most motor control study paradigms are paced not by movement duration, but rather by movement speed.
For such paradigms, movement duration has to be non-trivially estimated, or recorded from the data before modelling is even possible.
Furthermore, for both visual and physical perturbation paradigms the movement duration is further modulated by movement speed or perturbation timing \cite{Shadmehr1994b, Nashed2012a, Cross2019, Cesonis2020}. 
Our mixed-horizon controller allows us to bypass the requirement of movement duration all together, making this controller more human-like.
Second, the requirement of a movement time can be reliably bypassed via infinite-horizon control \cite{Qian2013, Jiang2011} or receding-horizon control \cite{Guigon2019} to generate controller movements with kinematics similar to experimental results.
However, such simulated movements do not contain any variations in the feedback gains throughout a movement (Figure \ref{fig:ofccomp}B and \cite{Cesonis2020}), while the presence of such feedback variations is foundational to numerous motor control studies \cite{Cesonis2020, Dimitriou2013, Franklin2008, Reichenbach2014, OostwoudWijdenes2011, Crevecoeur2013, Nashed2012a, Nashed2014, Zhang2018}.
Instead, the mixed-horizon OFC is able both select an appropriate movement duration and predict the temporal pattern of feedback responses.

We constructed the mixed-horizon OFC by employing the infinite-horizon controller and finite-horizon controller in series.
One of our key motivations for such a combination was that the finite-horizon controller needs movement durations as an input argument.
In particular cases, where only data-descriptive modelling is of interest, this is not a major technical issue -- simply extracting movement durations from the data for different types of movements, and using it with the finite-horizon OFC would produce identical behaviour to the behaviour of the mixed-horizon model.
However, the mixed-horizon control allows us to also generalise the model behaviour beyond only the available data-points to either make predictions about unseen conditions, for novel studies, or to provide some context to the available data.
The latter was particularly critical for our simulations of study 3 \cite{Franklin2016}, where we could simulate new experimental predictions even in the absence of experimental data, and therefore movement durations.
On the other hand, adjoining a finite-horizon controller to the infinite-horizon controller produces appropriate modulation of feedback responses throughout the movement that better resemble those of human participants.
While this approach may not be required if the only goal is to reproduce kinematics recorded in the data, it nevertheless produces a more human-like response.

Our mixed-horizon controller aims to combine the benefits of finite-horizon and infinite-horizon controllers, here focusing on perturbed, goal-directed movements.
The receding-horizon control \cite{Guigon2019} is another OFC implementation that combines the strengths of the infinite and finite-horizon control, although there are important differences between the two implementations.
The receding-horizon OFC uses a similar architecture as the finite-horizon OFC, however it neither needs, nor has any information about the movement duration prior, or even during the movement.
Instead, the movement is executed as a series of control signals with the horizon a fixed time away from the current state, until the end of the movement simply ``happens'' as the hand arrives at the goal.
Such implementation is very powerful when modelling long, slow movements, as these movements can be subdivided into movements with via-points, and consequently can accurately replicate the human behaviour in these situations.
However, for short, goal-directed movements without via-points the receding-horizon OFC could not replicate some features like visuomotor feedback responses in humans \cite{Cesonis2020}.
On the other hand, our mixed-horizon OFC combines the strengths of infinite and finite implementations into a different architecture, which continuously monitors the remaining time horizon for the movement and produces a control signal that matches that of humans. 
While for long or via-point movements the assumption that the whole time-horizon is predefined in advance may not be realistic, this is critically important to simulate the human-like behaviour in the fast, goal-directed reaching movements.

The combination of two separate control stages in series has previously been discussed in the neuroscience literature \cite{Wong2015, Cisek2010, Gallivan2016}.
In such a context the process of the infinite-horizon part of the control would be considered motor selection, whereas the finite-horizon part would be considered motor planning.
If the finite-horizon part is switched off, the controller is still capable of producing movements to the target, albeit showing less task-dependent modulation.
Similarly, human participants are also capable of reaching towards a target (with reduced accuracy) even if no motor planning is allowed \cite{Ghez1997}.
A more recent study has also associated the movement planning stage with trajectory calculation in obstacle avoidance tasks \cite{Wong2016}.
In that study, the authors demonstrated that a significant portion of reaction time after a go cue is spent calculating the movement trajectory if a specific trajectory was required, while these reaction times were reduced when no trajectory planning was necessary.
Here the finite-horizon controller is again consistent with the motor planning stage, as it adds the complexity to the movement features (in our case to movement dynamics).
While we have interpreted our two controllers as motor selection and motor planning, the literature also suggests other possible interpretations, e.g. motor planning and execution \cite{Gallivan2016, Kodl2011}.  

While motion planning and execution are generally considered as two processes in series, with planning occurring before execution, it is important to note that some kind of re-planning must occur in order to complete the task if the movement goal suddenly changes \cite{Georgopoulos1981}.
In our mixed-horizon implementation, the perturbation triggers the recalculation of the remaining movement duration (time-to-target) via the infinite-horizon OFC whenever it occurs.
This new duration is then used by the finite-horizon OFC to update the remaining control policy.
While these processes appear to be in series at any particular time point, both processes continually operate throughout the entire movement: the infinite-horizon is continuously waiting for any possible changes to trigger the re-planning and sending an updated time-to-target (or ``urgency'' signal \cite{Crevecoeur2013, Cesonis2020}) to the finite-horizon controller.
Importantly, both controllers still function with the same control weights, thus the only necessary stimulus from the environment to successfully complete the control is the error signal of the perturbation.
A similar idea has been previously described by \cite{Franklin2016b}, where fast visuomotor responses were explained via the pre-computation of feedback gains for possible perturbations.
Here, instead of pre-computing the responses to particular perturbations, the controller maintains fixed time-variable finite-horizon gains.
In turn, the infinite-horizon controller updates the time-to-target in case of any perturbations, which modulates the response of the finite-horizon, and thus overall the mixed-horizon controller. 
An alternative explanation is that both the infinite and finite controllers are continually re-calculated throughout the entire movement (rather than waiting for a discrete error signal), producing similar updates of the time-to-target or urgency after any perturbation.
While this would not change the model predictions in this paper, it is an open question whether the biological sensorimotor control system implements such continuous computation or re-computation only in the case of errors signals.

In this article we were interested not only in building the model and evaluating its dynamics, but also in how the model parameters and costs affect the outcome of the simulation.
Particularly, we analysed our mixed-horizon OFC sensitivity to individual optimisations for the same kinematic cost function, as well as sensitivity to different kinematic cost functions \ref{met:eqn5}.
Across multiple optimisations to the same cost function (optimisation sensitivity analysis) we observed the structure within controller costs $\omega_p$, $\omega_v$ and $\omega_r$ ($\omega_f$ was fixed to $\omega_v$, consistent with \cite{Liu2007}).
For the infinite-horizon part, outputs $\omega_p$, $\omega_v$ and $\omega_r$ of repetitive identical optimisations were distributed on the same plane perpendicular to the plane $\omega_p$-$\omega_v$ .
As such, instead of optimising for all three parameters simultaneously to obtain best fit controller costs, the problem can be reduced to a one-dimensional optimisation on $\omega_v$ for an arbitrary selected pair of ($\omega_p$,$\omega_r$).
Similarly, the finite-horizon part of controller outputs to the same repeated optimisation were generally distributed along a single line in the parameter space, which can also be reduced to a one-dimensional optimisation.
This is conceptually similar to the idea of structural learning \cite{Braun2009, Braun2010b}, where one meta-parameter could represent a movement along a task structure which is otherwise non-trivial if the entire parameter space needs to be searched.
Computationally it can be used to significantly reduce optimisation durations for similar problems by reducing the dimensionality of the problem.
That is, new movements may not require optimisation across a huge parameter space, but instead only optimisation along a single dimension which might easily be implemented during learning.

In addition to examining multiple optimisations to the same cost function, we also looked at the outcomes of optimisations with different kinematic cost functions.
We analysed this cost sensitivity in the context of our model for study 1 \cite{Cesonis2020} as it involves three different kinematics conditions and their corresponding dynamics.
Our mixed-horizon model provided mixed results when simulating the relative regulation of the response intensities between the conditions in our previous work \cite{Cesonis2020}.
Specifically, while the separable model could reliably simulate the upregulation and downregulation of the two conditions with changed kinematics, the non-separable model struggled to convincingly simulate the upregulation of the late-peak condition responses (Figure \ref{fig:St1_2}).
We suggest that this limitation arises, at least partially, from specific task requirements and their interactions with features of the OFC.
First, evidence from simulations suggests that contrary to humans, OFC simulates movements that are not exactly bell-shaped \cite{Cesonis2020, Berniker2019, Liu2007, Nashed2014} but instead have positively skewed asymmetric velocity profiles (peak velocities early in the movement).
This feature is particularly emphasised in the infinite-horizon problems \cite{Jiang2011, Qian2013, Cesonis2020}, as the movement costs are constant throughout the movement.
Furthermore, as we previously showed \cite{Cesonis2020}, we can not easily modulate the shape of this velocity profile for the infinite-horizon control, which means that our infinite-horizon controller is always projecting a movement with a velocity profile that peaks within the first half of the movement.
In turn, even if we fit the infinite-horizon controller to produce appropriate movement durations after perturbations, the difference in profile still introduces differences in perturbation timings, which then influences the responses. 
This is most strongly emphasised for the late-peak condition, as the separation between the finite-horizon and infinite-horizon kinematics is largest.
As a result, we posit that while this is indeed a model limitation, it minimally affects the movements that could be considered natural.

The cost sensitivity analysis of our models allows us to evaluate the general features of model behaviour.  
That is, we can separate model predictions that are outcomes of the parameter fitting from the predictions that are fundamental to the model (i.e. independent of the cost function or parameters).
Our results for study 1 \cite{Cesonis2020} and for sensitivity analysis (Figure \ref{fig:costsummary}) demonstrate consistent relative regulation of feedback responses across the three experimental conditions, with early-peak condition down-regulated and late-peak condition up-regulated from the baseline.
Importantly, these results are consistent throughout our vast set of explored kinematic cost functions that result in an even broader set of model parameters.
Furthermore, our model has no further assumptions beyond the LQG-governed optimal feedback controller, which is enough to replicate the behaviour of human participants.
This suggests that the behaviour observed in the data could be entirely governed by the response of the optimal control policy to the movement requirements, and does not require any additional control components. 

Our results for study 2 replicated the main behaviour of experiment 2 from Dimitriou and colleagues \cite{Dimitriou2013}, showing that combined cursor and target perturbations produce weaker responses than those to an isolated perturbation (Figure \ref{fig:dim_sep}).
We obtained this result for both separable and non-separable models.
While the original authors discussed this down-regulation as arising due to the uncertainty of the limb, our models included no assumptions about the certainty and thus such behaviour is a direct outcome of the OFC.
In contrast, we were not able to replicate the behavioural results of experiment 3 \cite{Dimitriou2013}, where the feedback gains were conditionally modified (either up- or down-regulated) when the target perturbation preceded the cursor perturbation. 
Instead, our model continued to produce similar down-regulation for combined perturbations as in experiment 2.
The original authors suggested that the experiment 3 results may have occurred from a recalculation of the control gains during the 100 ms between the target and cursor perturbations. 
However, our model implementation immediately recalculates optimal control gains after any perturbations by default (including in the model for experiment 2).
Here, based on the model-data discrepancies, we presume that instead of just recalculating the control policy after the target is perturbed along the movement direction, human participants could have also updated the kinematic cost function in a non-trivial way that generated such behaviour.

Our simulations of study 3 \cite{Franklin2016} demonstrate the applicability of the mixed-horizon OFC in situations where data is not currently available. 
Specifically, in the original study only a single perturbation location was used to measure the visuomotor feedback responses across a range of conditions.  
One particular observation was a non-linear increase in feedback responses with perturbation size (Figure \ref{fig:dfseparable}A), which was not explained. 
Here we examined whether our mixed-horizon optimal controller could explain this non-linear increase in feedback intensities by simulating more data points than experimentally collected.
In turn, we demonstrated that such scaling is a combination of a linear gain regulation with perturbation size further modulated by the extension in movement duration, which also depends on perturbation size.
It is important to note that we would not have been able to generate such results with only finite, infinite or receding horizon models, either due to the lack of input data (movement durations for not recorded perturbation locations in finite-horizon simulations), or due to model's inability to simulate the temporal evolution of feedback gains (infinite and receding horizon).

Our proposed mixed-horizon model is able to both predict movement durations for untrained conditions, and generate human-like modulation of feedback intensities throughout movements, by combining features of both the infinite and finite-horizon controllers.
However, one limitation in combining these two separate models, as currently implemented, is that each model has their individual control parameter sets.
While in the ideal case the two parts would share parameters, this is not possible due to the mathematical implementations of infinite and finite-horizon OFC.
Finite-horizon OFC models utilise cost functions Q and R that are (or at least can be) variable over time, while infinite-horizon models use stationary costs.
As both of these models serve different purposes within the whole mixed-horizon control, such architecture implies that the infinite and finite-horizon models are two separate systems within the overall control, rather than one holistic system.
Despite these limitations, the mixed-horizon OFC is able to model a large range of behaviours and paradigms that were not previously possible, as well as make predictions for new studies.

In this paper we have demonstrated a new mixed-horizon approach to optimal feedback control modelling of human behaviour.
Specifically, we have contrasted the results of the finite and infinite-horizon simulations.
While each implementation is valuable on its own, it also has limitations in modelling specific features of human motor control, such as the temporal evolution of feedback gains or variable movement duration of perturbed movements.
By combining the infinite-horizon and finite-horizon controllers into a mixed-horizon controller we were able to better model the results of previously published studies and provide alternative explanations of their results \cite{Dimitriou2013}, further reinforce the results \cite{Cesonis2020}, or model the results that could not previously be simulated \cite{Franklin2016}.
{All together our results demonstrate a novel and powerful approach to optimal feedback control models that can more accurately represent behaviours observed in humans.}

\section*{acknowledgements}
We thank Clara Günter and Raz Leib for their comments on this manuscript.

\bibliography{sample_bib}

\end{document}